\documentclass[fleqn,usenatbib]{mnras}

\usepackage{newtxtext,newtxmath}

\usepackage[T1]{fontenc}

\DeclareRobustCommand{\VAN}[3]{#2}
\let\VANthebibliography\thebibliography
\def\thebibliography{\DeclareRobustCommand{\VAN}[3]{##3}\VANthebibliography}

\usepackage{graphicx}	
\usepackage{amsmath}	
\usepackage{siunitx}
\usepackage{xspace}
\usepackage{booktabs}
\usepackage{tabularx}
\usepackage{hyperref}
\usepackage{multirow}
\usepackage{orcidlink}

\usepackage[dvipsnames]{xcolor}
\newcommand{\Multi}{\texttt{MULTI}\xspace}
\newcommand{\MultiD}{\texttt{MULTI3D}\xspace}
\newcommand{\Dispatch}{\texttt{DISPATCH}\xspace}
\newcommand{\Must}{\texttt{M3DIS}\xspace}
\newcommand{\Turb}{\texttt{TURBOSPECTRUM}\xspace}

\newcommand{\Stagger}{\texttt{STAGGER}\xspace}
\newcommand{\Cobold}{\texttt{CO5BOLD}\xspace}
\newcommand{\Marcs}{\texttt{MARCS}\xspace}

\DeclareSIUnit\angstrom{\text {Å}}

\title[The solar carbon abundance from CH lines.]{Solar carbon abundance from 3D non-LTE modelling of the diagnostic lines of the CH molecule}

\author[R. Hoppe et al.]{
Richard Hoppe$^{1,2}$\orcidlink{0000-0002-8451-6260},\thanks{E-mail: hoppe@mpia-hd.mpg.de.de}
Maria Bergemann$^{1}$\orcidlink{0000-0002-9908-5571},
Philipp Eitner$^{1,2}$\orcidlink{0009-0007-4502-8081},
Momo Ellwarth$^{3,4,5}$\orcidlink{0009-0007-0922-7315},
\AA{}ke Nordlund$^{6}$\orcidlink{0000-0002-2219-0541},
\newauthor
Jorrit Leenaarts$^{7}$\orcidlink{0000-0003-4936-4211},
Bertrand Plez$^{8}$\orcidlink{0000-0002-0398-4434},
Aldo Serenelli$^{9}$\orcidlink{0000-0001-6359-2769},
\\
$^{1}$ Max-Planck-Institut für Astronomie, 69117 Heidelberg, Germany\\
$^{2}$ Ruprecht-Karls Universität, 69117 Heidelberg, Germany\\
$^{3}$ Institut für Astrophysik und Geophysik, 37077 Göttingen, Germany\\
$^{4}$ Lowell Observatory, 1400 W. Mars Hill Road, Flagstaff, AZ 86001, USA\\
$^{5}$ Department of Astronomy and Planetary Science, Northern Arizona University, PO Box 6010, Flagstaff, AZ 86011 USA\\
$^{6}$ University of Copenhagen, Niels Bohr Institute, Copenhagen, DK-2100, Denmark \\
$^{7}$ Institute for Solar Physics, Department of Astronomy, Stockholm University, AlbaNova University Centre, SE-106 91 Stockholm, Sweden \\
$^{8}$ Laboratoire Univers et Particules de Montpellier, Univ Montpellier, CNRS, Montpellier, France \\
$^{9}$ Institute of Space Sciences (ICE-CSIC), Barcelona, Spain
}

\date{Accepted 2025 November 17. Received 2025 October 16; in original form 2025 May 22}

\pubyear{\the\year{}}

\begin{document}
\label{firstpage}
\pagerange{\pageref{firstpage}--\pageref{lastpage}}
\maketitle

\begin{abstract}
\textit{Context.}
The spectral lines of the CH molecule are a key carbon (C) abundance diagnostic in FGKM-type stars. These lines are detectable in metal-rich and, in contrast to atomic C lines, also in metal-poor late-type stars.  However, only 3D LTE analyses of the CH lines have been performed so far.

\noindent\textit{Aims.}
We test the formation of CH lines in the solar spectrum, using for the first time, 3D Non-LTE (NLTE) models. We also aim to derive the solar photospheric abundance of C, using the diagnostic transitions in the optical (\SI{4218}{} - \SI{4356}{\angstrom}) and infrared (\SI{33025}{} - \SI{37944}{\angstrom}).

\noindent\textit{Methods.}
We use the updated NLTE model molecule from \cite{PopaHoppe2023} and different solar 3D radiation-hydrodynamics model atmospheres. The models are contrasted against new spatially-resolved optical solar spectra, and the center-to-limb variation (CLV) of CH lines is studied.

\noindent\textit{Results.}
We find generally small ($\sim$0.01 dex) NLTE effects in the optical and IR diagnostic CH A-X lines in the solar atmosphere. Both 3D NLTE and 3D LTE spectral modelling yield an excellent fit to the solar intensity observations at all viewing angles.
The 1D LTE and 1D NLTE models fail to describe the line CLV, and lead to underestimated solar C abundances. The 3D NLTE modelling of diagnostic lines in the optical and IR yields a carbon abundance of A(C)=$8.52\pm0.07$ dex. The estimate is in agreement with recent results based on neutrino fluxes measured by Borexino.

\noindent\textit{Conclusions.}
3D NLTE modelling and tests on spatially-resolved solar data are essential to derive robust solar abundances. The analysis presented here focuses on CH, but we expect that similar effects will be present for other molecules of astrophysical interest.
\end{abstract}




\section{Introduction}
The CH molecule, with its rich spectral features, is an invaluable tool for tracing carbon (C) abundances in stellar atmospheres. Starting from early studies like \cite{Russell1934}, the CH spectral lines, especially the critical G-band of CH at \SI{4300}{\angstrom}, were used to determine C abundances in metal-poor stars, red giants, stellar populations \citep{ShetroneSmith1999, YoonBeers2016, YoonBeers2018, LucatelloBeers2006, StarkenburgHill2013, PlaccoKennedy2011, HansenNordstrom2016, Suarez-AndresIsraelian2017}.

The G-band is particularly important in spectra of metal-poor stars, where atomic carbon (C I) lines are weak and difficult to disentangle from other features \citep[e.g.][]{WheelerSneden1989, AlexeevaMashonkina2015}.
The solar C abundance is important in the calculation of Standard Solar Models (SSM), which describe the structure of the solar interior and its evolution \cite[e.g.][]{SerenelliBasu2009, PinsonneaultDelahaye2009, VinyolesSerenelli2017}.

In the light of large spectroscopic surveys like GALAH, 4MOST and WEAVE, it is necessary to establish a robust reference for the analysis of stellar C abundances based on stellar spectra. So far all spectroscopic studies have resorted to 1D LTE or 3D LTE modelling of CH lines in stellar spectra \citep{Sneden1974, CarbonBarbuy1987, TomkinLemke1992, SpiteCayrel2005, ColletAsplund2007, BonifacioSpite2009, BeharaBonifacio2010, HayekAsplund2011, BonifacioCaffau2018}. In particular, it has been shown that the effects of convection are crucial, and lead to systematic increase of molecular band strengths in metal-poor stars, [Fe/H]$\lesssim -2$ \citep[e.g.][]{AsplundGarciaPerez2001, ColletAsplund2006, ColletAsplund2007, HayekAsplund2011, ColletNordlund2018, FrebelCollet2008, GallagherCaffau2016, NorrisYong2019}. This is due to adiabatic cooling associated with convection in realistic 3D radiation-hydrodynamics (RHD) models. The effects of convection on abundances are of the order $-0.2$ dex to $-0.5$ dex at [Fe/H] $\lesssim -2$  \citep{GallagherCaffau2016}. 

Also for the Sun, detailed spectral analysis of CH lines in 1D LTE and 3D LTE was performed. The 1D LTE studies include \cite{GrevesseLambert1991, AlexeevaMashonkina2015}, whereas 3D LTE studies are \cite{AsplundGrevesse2005, AsplundGrevesse2009, AsplundAmarsi2021}. The 3D LTE studies found qualitatively different effects of 3D convection in the analysis of the diagnostic CH lines in the optical and infra-red wavelength ranges. Whereas \citep{AsplundGrevesse2005} found that the 3D LTE value of A(C) is within 0.01 dex of the 1D LTE result, the most recent analysis \citet{AsplundAmarsi2021} found a higher 3D LTE A(C) compared to the value obtained with a 1D \Marcs models by 0.03-0.07 dex.  However, the effects of departures from LTE have not been explored in 3D in any detailed systematic approach to date.

In the NLTE framework, interactions between the radiation field and the gas in stellar atmospheres are calculated by solving the coupled equations of radiation transfer and statistical equilibrium using realistic cross-sections and reaction rates from the lab or theoretical studies \citep[e.g.][]{ Asplund2005, HubenyMihalas2014, BergemannNordlander2014, LindAmarsi2024}. These are vastly more complex compared to the very simple LTE approximation, which assumes the state of maximum entropy. Studies investigating NLTE effects in molecules have been scarce and focussed on CO \citep{AyresWiedemann1989, SchweitzerHauschildt2000, BerknerSchweitzer2015} and H$_2$O \citep{LambertJosselin2013}, but also with highly simplified molecular models, ignoring detailed key reactions like state-resolved photo-dissociation cross-sections or charge exchange reactions. Recently, NLTE effects in the CH diagnostic spectral lines were studied with 1D stellar atmosphere models in our recent work \cite{PopaHoppe2023} (hereafter PH23).
 
In this paper, we provide a systematic analysis of the solar C abundance based on molecular CH lines in the solar spectrum. Different from the previous work, we use NLTE modelling with 3D RHD simulations of solar convection, and we contrast the predictions of models against new spatially-resolved observations of the optical wavelength range \citep{EllwarthSchafer2023, ReinersYan2023}. These center-to-limb spectra (\SI{4200}{} to \SI{8000}{\angstrom}) cover the optical range of the CH G-band, allowing an entirely novel test of the solar spectroscopic models. We explore the consistency of C abundances retrieved from spectra taken at different angles, and explore the physics underlying 1D LTE, 1D NLTE, 3D LTE and 3D NLTE line formation of the diagnostic CH lines. Finally, we provide the solar 3D NLTE C abundances based on optical and IR lines, and analyse the new value in the context of other measurements and probes, including the solar wind \citep{vonSteigerZurbuchen2016} and data obtained from the Borexino particle physics experiment \citep{Gonzalez-GarciaMaltoni2024}.

\begin{figure}
    \centering
    \includegraphics[width=\linewidth]{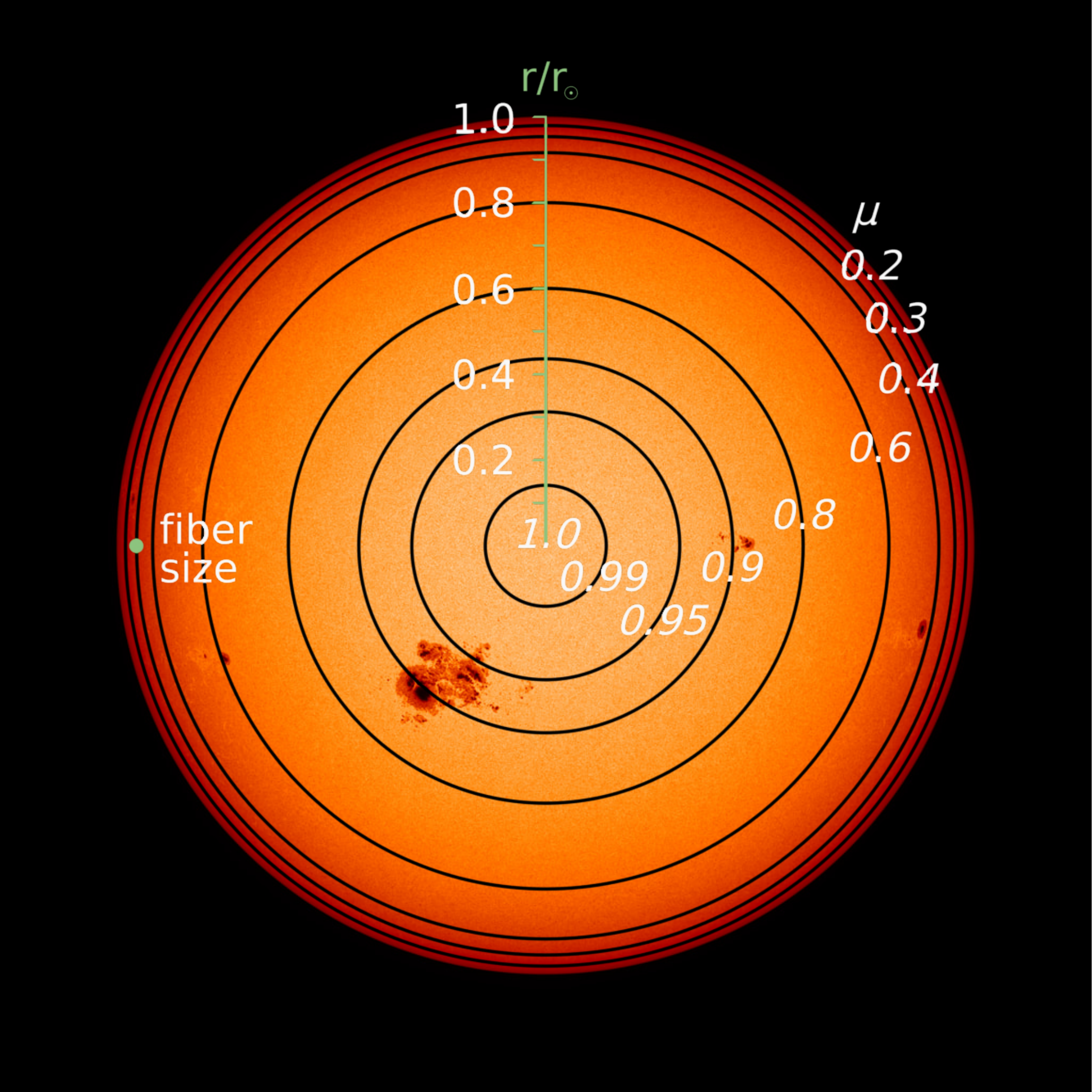}
    \caption{A view of the solar disk by the Helioseismic and Magnetic Imager \protect\citep[image by NASA's Scientific Visualization Studio, ][]{nasa_sun_image}. Overlayed are concentric rings showing viewing positions of constant $\mu$-angle, while the green scale shows corresponding relative radii. We also show the finite fiber size of \SI{32.5}{\arcsec} of the IAG instrument \protect\citep{SchaferRoyen2020}.}
    \label{fig:SunMu}
\end{figure}

\begin{figure}
    \centering
    \includegraphics[width=\linewidth]{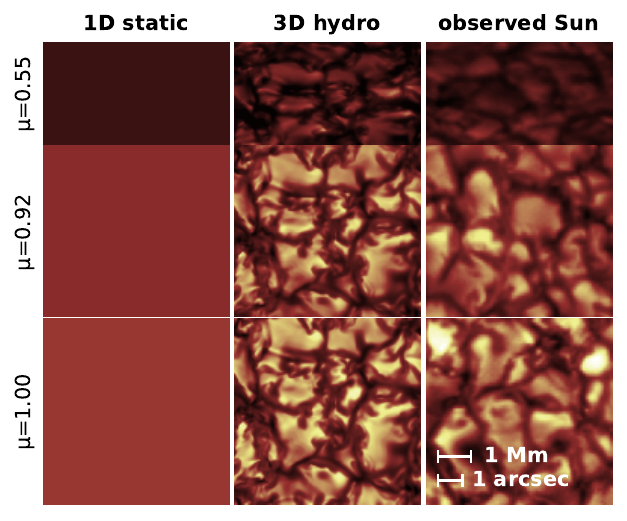}
    \caption{Comparison of synthetic and observed solar G-band intensity at wavelength $4306.8\pm$\protect\SI{15}{\angstrom} at different positions across the solar disc (different rows). The synthetic G-band has been convolved with the SST photometric filter transmission function \protect\citep{CarlssonStein2004}.
    The G-band observations were acquired with the Swedish 1-m Solar Telescope (SST, \protect\cite{Scharmer2003a}) on La Palma in 2003-2004 \protect\citep[][Van Der Voort priv. comm.]{Berger2004, Lin2005}. For more detailed information on these observations we refer the reader to \url{https://www.isf.astro.su.se/data1/gallery/}. We show the disc center $\mu=1$ at the bottom and two angles approaching the limb $\mu=0.92, \mu=0.55$ above. The black and white levels of the observations were adjusted to account for the different processing of the individual measurements.
    }
    \label{fig:SunVoort}
\end{figure}

\section{Analysis}
\subsection{Observational data}
The observed spatially-resolved spectroscopic data around \SI{4300}{\angstrom} are taken from the  solar intensity FTS atlas by \cite{EllwarthSchafer2023} taken at the Institute for Astrophysics and Geophysics(IAG), G\"{o}ttingen. The atlas covers the full wavelength range between \SI{4200} - \SI{8000}{\angstrom} and contains observations of the solar disk at 14 different viewing angles between $\mu=1.0$ and $\mu=0.2$. The $\mu$-angle and relative radius $r/r_\odot$ are defined as
\begin{align}
    \mu &= \cos\theta \\
    \frac{r}{r_\odot} &= \sin\theta = \sqrt{1 - \mu^2},
\end{align}
where $\theta$ is the angle between the line-of-sight and the normal of the solar surface (see Fig. \ref{fig:SunMu}). 

The resolving power of the observations is $R=\nu/\Delta\nu\sim$\SI{1\,000\,000}{} while the signal-to-noise (SNR) increases with wavelength. Due to the comparatively poor SNR of 10-50 at the \SI{4000}{\angstrom} end we have decided to exclude those angles where there were less than 10 individual IAG observations, leaving us with the 9 angles $\mu=$ 1.0, 0.99, 0.95, 0.9, 0.8, 0.6, 0.4, 0.3, 0.2.

For the infrared transitions at ($32\,946-$\SI{37955}{\angstrom}) we use the Spacelab-3 \texttt{ATMOS} atlas ($\mu=0.935$) \citep{FarmerNorton1989} as uploaded by \cite{SeoKim2007} (file \texttt{solspec2b\_1.txt}) \footnote{\href{https://www.sciencedirect.com/science/article/pii/S0019103507003089?via\%3Dihub\#ec005}{\nolinkurl{https://www.sciencedirect.com/science/article/pii/S0019103507003089?via\%3Dihub\#ec005}}} as NASA's original links are deprecated. Due to wiggles in the continuum we renormalise the atlas utilising the neural network normalisation tool \texttt{SUPPNet} \citep{RozanskiNiemczura2022} with the sampling set to \SI{0.05}{\angstrom}, a smoothing factor of $2$ and no further manual adjustment of the knots of the continuum spline.

The G-band observations in Fig. \ref{fig:SunVoort} were acquired with the Swedish 1-m Solar Telescope (SST, \cite{Scharmer2003a}) on La Palma in 2003-2004. The images were acquired with an interference filter centred on \SI{4305}{\angstrom} and with bandwidth of \SI{11}{\angstrom}. Seeing conditions were such that near-diffraction limited imaging was achieved with the aid of adaptive optics \citep{Scharmer2003b} and image restoration \citep{Löfdahl2002, VanNoort2005}. The $\mu=1$ and the $\mu=0.92$ observations were published in \cite{Berger2004} and \cite{Lin2005}. 

\subsection{The Code}
We use our updated version of the 3D NLTE radiative transfer code \MultiD within the \Dispatch framework, which provides a task-based parallelisation scheme through MPI as well as openMP  \citep{NordlundRamsey2018}.

Even though the original version of \MultiD \citep{LeenaartsCarlsson2009} already featured spatial and frequency domain decomposition, the task-based approach of the \Dispatch framework has several advantages which will be discussed in an accompanying paper (Hoppe in prep.). Besides the new parallelisation scheme we have updated the background bound-free opacities \citep[see][]{EitnerBergemann2024}. We are confident that the radiative transfer, EOS, background opacities and broadening recipes work reliably, as we tested synthetic CH line profiles computed with \MultiD against the independent spectrum synthesis code \Turb in 1D LTE (see Fig. \ref{fig:TScomparison} in Appendix).

In the NLTE calculations for the CH molecule, we included background bound-bound opacities during the statistical equilibrium calculations via interpolation of opacity tables precomputed from \Marcs \citep{GustafssonEdvardsson2008}.

For the purposes of this work, we employ \textit{auto-preconditioning} \citep[][Section 2.4]{RybickiHummer1992}, which suppresses only the feedback in the radiative rates of each line with itself. This approach is also used in other 3D NLTE codes such as Balder \citep{AmarsiBarklem2019}. In our case, it is adopted to reduce the computational overhead, which becomes significant when dealing with a large number of energy levels in the \textit{full-preconditioning} scheme \citep[][Section 2.3]{RybickiHummer1992} originally implemented in \MultiD.

\subsection{Diagnostic CH lines used for the solar C abundance analysis}

\begin{figure}
    \centering
    \includegraphics[width=\linewidth]{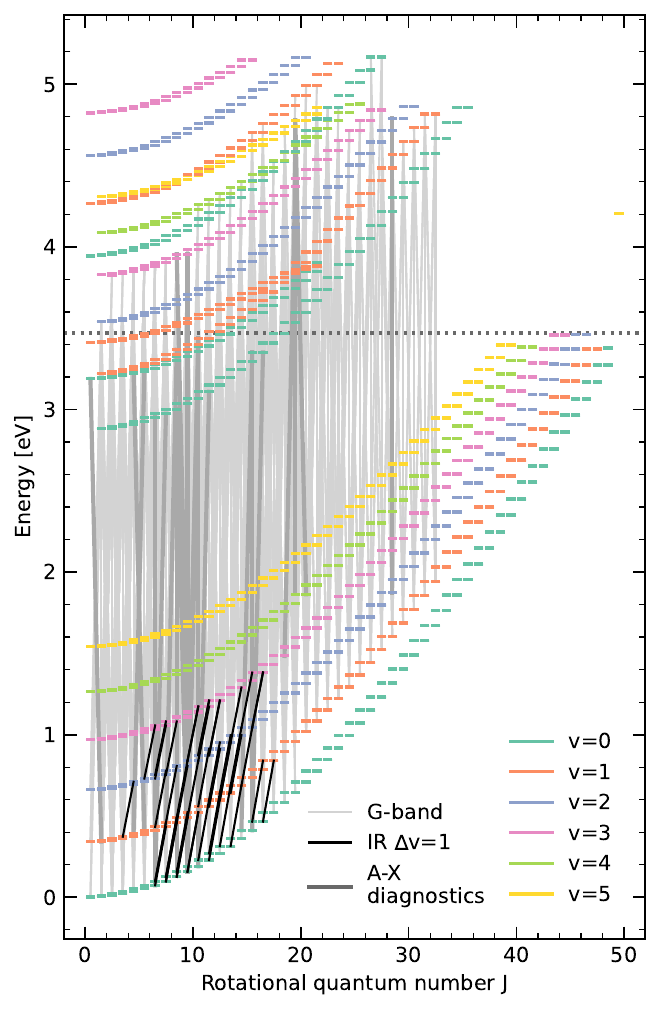}
    \caption{Grotrian diagram of the CH molecule.}
    \label{fig:Grotrian}
\end{figure}

Our line selection is based on the one by \cite{AsplundGrevesse2005} (AGS05) who focused on 9 optical A-X transitions and 102 infrared rotation-vibration transitions to determine the solar C abundance. While we adopted the same 9 optical transitions (table \ref{tab:AX_data}), we reduced the infrared line sample down to 50 lines (table \ref{tab:IR_data}). We removed lines from the infrared sample where the continuum normalisation by SUPPNet was unsatisfactory or the line was positioned in the wing of a stronger line after visual inspection. We also removed those lines resulting from superposition of multiple CH transitions being spaced less than \SI{0.7}{\angstrom} apart. This was done because the theoretical line centers from the \cite{MasseronPlez2014} line list do not perfectly match the observed line centers to the \SI{0.1}{\angstrom} level. For single lines this can be accounted for by slightly shifting the line in the fit. However, when the spacing between the centers of two blended lines is off, the compound shape of the lines does not necessarily fit the observed shape. All diagnostic transitions are highlighted in the Grotrian diagram of the CH molecule in Fig. \ref{fig:Grotrian} showing its different electronic, vibrational and rotational enery states.

\subsection{The CH molecule model}
We adopt the CH molecule model employed in PH23, which consists of \SI{1981} energy states, \SI{18377}{} bound-bound transitions between \SI{1000}{} - \SI{20000}{\angstrom} with line data taken from \cite{MasseronPlez2014} and \SI{932}{} photo-dissociation transitions based on data by \cite{KuruczvanDishoeck1987}. Any modifications to the original model employed in PH23 are discussed in \ref{sec:updates}. We employ the partition function data from \cite{SauvalTatum1984} for the CH molecule. The molecular dissociation energies employed in our code match the ones from the more recent compilation of \cite{BarklemCollet2016}. We compared the partition functions of CH and CO from \cite{SauvalTatum1984} to the partition functions from \cite{BarklemCollet2016} and found that there is agreement at the sub-percent level for the partition function of CO in the temperature range of \SI{1000}{\K}-\SI{10000}{\K} between the two datasets. However, for CH \cite{BarklemCollet2016} predict a higher partition function by about 5-$7\%$ compared to \cite{SauvalTatum1984} in the same temperature range. 
To test the effect, we adopted the partition functions from \cite{BarklemCollet2016} in our code. We find negligible differences in the derived C abundances in 1D LTE below the 0.01 dex level. This is because the partition function enters the level population calculation as well as the chemical equilibrium calculations. An increased CH partition function results in a larger amount of C being locked in CH, but reduces the individual fractional level populations. Therefore our results are not affected by the choice of the partition function.

\subsubsection{Superlevels}
In order to perform 3D NLTE calculations, the model molecule had to be reduced, which is a common procedure for atoms \citep{BergemannLind2012}. This is because it is in practice extremely time-consuming to do matrix inversions for a over a few hundred energy states coupled by radiation and collisional transitions.

Therefore, we combined energy levels that are very close in excitation energy value into superlevels \citep[e.g.,][]{HubenyMihalas2014}. For this we used a K-means clustering algorithm in Python to group and merge the levels with similar excitation energy. Besides merging energy levels we also took great care to merge photo-dissociation transitions accordingly. After reducing the CH molecule to $294$ energy levels the computational cost of the matrix inversions was negligible, while the non-LTE effects on the populations were preserved, as discussed in section \ref{sec:reduced} in the Appendix.

\subsubsection{Bound-bound transitions}
We did not merge any bound-bound transitions, in order to preserve the Voigt profiles of each single transition. For this we had to overcome a limitation of the code, as by default it determines the central wavelength of each line from the energy difference between the upper and lower state of the transition. Instead we introduced an alternative model atom/molecule format, which tabulates exact line frequencies and log(gf) values for each transition, very similar to common line list formats. Each spectral line is still associated with an upper and lower state from the reduced molecule. When dropping the assumption of LTE, we assume that upper and lower level departure coefficients are given by the departure coefficients of the respective superlevels. Hence, we assume that levels with similar excitation energies have equal departure coefficients.

\subsubsection{Collisional transitions}
Excitation and dissociation processes of the CH molecule have been studied extensively in the context of combustion and interstellar chemistry, but little in the context of stellar atmospheres. The most important collision partners in astrophysical environments are generally electrons, because of their high speed and frequent interaction rates, and H, He and H$_2$ as these are usually the most abundant species. Excitation of CH by collisions with electrons is generally considered insignificant as electron collisions only dominate at temperatures too hot for CH to be abundant.

Theoretical excitation rates have been presented for CH collisions with He \citep{MarinakisDean2015} as well as for collisions with H and H$_2$ \citep{Dagdigian2016, Dagdigian2017, Dagdigian2018}. These studies have probed the temperature range between 10-\SI{300}{\K} relevant for excitation of CH molecules in the interstellar medium. The behaviour of the collision rates as a function of temperature differs drastically between the different colliders. While collisions with H and He show a strong temperature dependence, rate constants for collisions with H$_2$ are approximately constant with temperature. The extrapolation to temperatures in stellar atmospheres has been discouraged by the authors \citep[see][]{PopaHoppe2023}.
More recently, empirical recipes for collisions between CH and He as well as H$_2$ have been presented by \cite{YanWang2025} based on a compilation of experimental collision data for NO, OH and CH molecules with a range of different colliders. This study focuses on the forward modelling of laser-induced-fluorescence (LIF) which aids the analysis of experimental LIF measurements, for example in combustion systems. While their recipes are applicable in a temperature range of 200-\SI{3000}{\K} , many of the recipe parameters for CH have been deduced by analogy from NO or OH.

A single measurement exists for the reaction rate of CH($v$=0, $j$=0) + H $\rightarrow$ C + H$_2$, obtained through a shock-tube experiment on its inverse reaction, C + H$_2$ $\rightarrow$ CH + H \citep{DeanDavidson1991}. The measured rate was $1.83\times10^{-10}$ \SI{}{\cm^3\per\s}, with an uncertainty of 30\%, over a temperature range of 1500–\SI{2250}{\K}  \citep[Table 3 in][]{ZhaoZhang2022}. Due to the lack of experimental and theoretical studies of CH collision rates, we rely on the Drawin recipe for excitation by collisions with atomic hydrogen. For collisional dissociation, we compared two approaches:  \\
1. Applying the Drawin recipe only to levels below the dissociation threshold.  \\
2. Using the experimental dissociation rate for all energy states in our CH model.  \\
Both methods produced abundance corrections of approximately +0.01 dex for the diagnostic lines in 1D NLTE calculations. Consequently, we adopted only the model incorporating Drawin collisions for our 3D NLTE calculations.

\subsection{Model atmospheres}

\begin{figure}
    \centering
    \includegraphics[width=\linewidth]{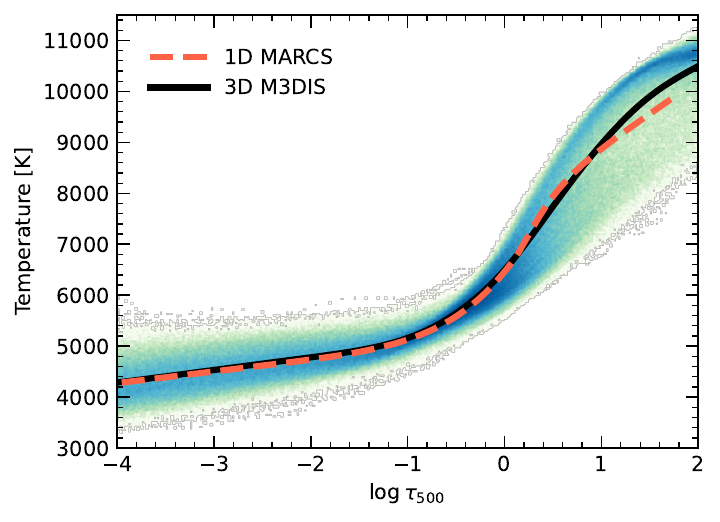}
    \caption{Temperature structure as a function of optical depth of employed solar models. All 11 downsampled \Must snapshot are presented as a density plot in the background. The solid black line shows the average of the snapshots over surfaces of equal $\tau_{500}$. Overplotted as red dashed line is the corresponding structure of the 1D \Marcs model.}
    \label{fig:Sun_temp}
\end{figure}

We compare the results from hydrostatic and 3D RHD atmospheric models of the Sun. We employ 11 snapshots from a solar \Must simulation \citep{EitnerBergemann2024} and 13 snapshots from a solar \Stagger simulation \citep{MagicCollet2013}, both of which are 3D RHD simulations. It has been demonstrated that sampling ten statistically independent snapshots is sufficient to achieve accurate line-profile fits, with a maximum flux error of 0.005 \citep{RodriguezDiazLagae2024}. Earlier solar calculations already showed that using as few as five snapshots yields reliable results \citep{BergemannGallagher2019, GallagherBergemann2020}. For comparison, we use a 1D hydrostatic \Marcs solar model \citep{GustafssonEdvardsson2008} with a microturbulence of $v_{\rm mic} = \SI{1}{\km\per\s}$. Additionally, we use two \textit{average 3D} (or <3D>) models, which are 1D models derived by averaging 3D hydrodynamical snapshots over surfaces of equal optical depth at \SI{5000}{\angstrom}. These are the <\Must> and <\Cobold> models. When using the <3D> atmospheres, we also assume a $v_{\rm mic} = \SI{1}{\km\per\s}$. We did not consider the \Cobold model in full 3D, as only the <3D> version of the model is available to us \citep{MaggBergemann2022}.
For this study, we resample both hydrostatic and hydrodynamic atmospheres based on their electron pressure gradient to ensure detailed sampling in the photosphere. This resampling has negligible impact on the spectra of 1D models, which are usually tabulated in equal optical depth steps, but is common practice in the case of hydrodynamical atmospheres which have to extend to optically thick layers $\log\tau \gg 1$ at the expense of fine sampling in the photosphere \citep{AmarsiNordlander2018, RodriguezDiazLagae2024}. The temperature structure of the \Must simulation is presented in Fig. \ref{fig:Sun_temp}. Average temperature differences between the rest of the employed atmospheres and how they affect the CH abundance are shown in Fig. \ref{fig:atmos_comparison} in the Appendix.

The 1D atmospheres are interpolated to have 128 points in total, whereas the 3D snapshots are interpolated to have $30\times 30 \times 128$ points each. This constitutes a significant downsampling of the 3D snapshots from their original resolution, but the impact on the derived abundances is below $0.01$ dex in the diagnostics lines (as shown in Fig. \ref{fig:resolution} in the Appendix).

\begin{figure}
    \centering
    \includegraphics[width=\linewidth]{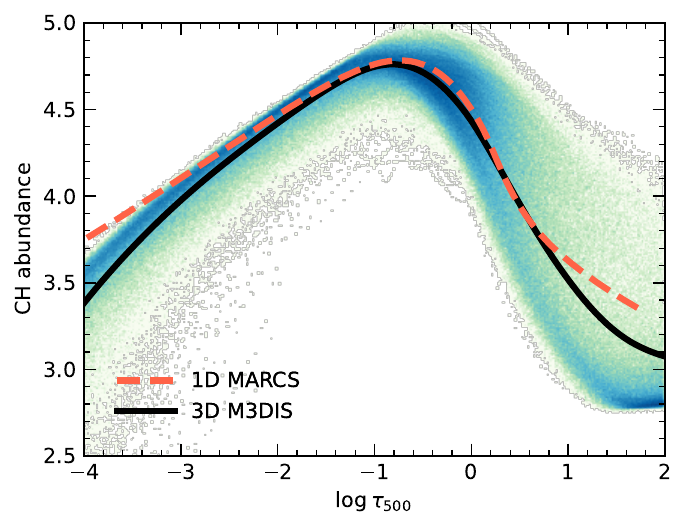}
    \caption{CH LTE number densities converted to absolute CH abundance, $\log_{10}(N_{\rm CH}/N_{\rm H}) + 12$, as a function of optical depth. All 11 downsampled \Must snapshots are presented as a density plot in the background. The solid black line shows the average of all 3D snapshots over surfaces of equal $\tau_{500}$. Overplotted as red dashed line is the corresponding structure of the 1D \Marcs model.}
    \label{fig:Sun_abund}
\end{figure}

\subsection{Spectrum Synthesis}
We compute spectra for three different carbon abundances spaced by $0.4$ dex centred around $8.43$. Non-LTE spectra share a single set of departure coefficients, determined at the central abundance. For synthesis of spectra at intermediate abundances we use monotonic Hermite interpolation between the three reference spectra. We tested the impact of this interpolation method in 1D NLTE against computing departure coefficients and spectra directly with a given carbon abundance and found that it does not exceed $0.01$ dex in the abundance range between $8.35$ and $8.83$ dex for both optical and infrared lines (see Fig. \ref{fig:Interpolation} in the Appendix). \cite{ColletAsplund2007} showed that, in contrast to more metal-poor stars, the formation of CH is independent of the oxygen abundance in the Sun.

\section{Results}

\begin{table}
\centering
\caption{A-X line data employed in the CLV analysis. Most of the 9 discussed diagnostic transitions have multiple components. $E_i$ represents the lower excitation potential of each transition. $W_{\rm eq}$ represents the equivalent widths measured for each feature (multiple blended components) in the KPNO solar flux atlas.}
\label{tab:AX_data}
\begin{tabularx}{\linewidth}{
>{\centering\arraybackslash}X
>{\centering\arraybackslash}X
>{\centering\arraybackslash}X@{}
>{\centering\arraybackslash}X@{}
>{\centering\arraybackslash}X@{}
>{\centering\arraybackslash}X@{}
>{\centering\arraybackslash}X
>{\centering\arraybackslash}X}
\toprule
$\lambda$ [\AA] & $\log(g\!f)$ & $E_i$ [eV] & $\nu_i$ & $J_i$ & $\nu_j$ & $J_j$ &  $W_{\rm eq}$ [m\AA] \\
\midrule
4218.713 & -1.315 & 0.413 & 0 & 15.5 & 0 & 16.5 & \multirow{4}{*}{84.7} \\
4218.735 & -1.337 & 0.413 & 0 & 14.5 & 0 & 15.5 & \\
4218.744 & -4.377 & 1.484 & 3 & 18.5 & 3 & 19.5 & \\
4218.783 & -3.410 & 0.413 & 0 & 15.5 & 0 & 15.5 & \\
\noalign{\vskip 2mm}  
4248.873 & -3.809 & 1.874 & 2 & 28.5 & 2 & 28.5 & \multirow{3}{*}{73.0} \\
4248.938 & -1.431 & 0.191 & 0 & 10.5 & 0 & 11.5 & \\
4248.952 & -3.257 & 0.191 & 0 & 10.5 & 0 & 10.5 & \\
\noalign{\vskip 2mm}  
4252.937 & -4.346 & 1.863 & 4 & 20.5 & 5 & 19.5 & \multirow{3}{*}{51.0} \\
4252.955 & -1.814 & 1.863 & 4 & 19.5 & 5 & 19.5 & \\
4253.000 & -1.506 & 0.523 & 1 &  9.5 & 1 & 10.5 & \\
\noalign{\vskip 2mm}  
4253.204 & -1.471 & 0.523 & 1 & 10.5 & 1 & 11.5 & \multirow{2}{*}{46.5} \\
4253.216 & -3.279 & 0.523 & 1 & 10.5 & 1 & 10.5 & \\
\noalign{\vskip 2mm}  
4255.248 & -3.211 & 0.157 & 0 & 9.5 & 0 & 9.5 & \multirow{2}{*}{75.0} \\
4255.248 & -1.460 & 0.157 & 0 & 9.5 & 0 & 9.5 & \\
\noalign{\vskip 2mm}  
4263.973 & -1.575 & 0.460 & 1 & 7.5 & 1 & 8.5 & 47.0 \\
\noalign{\vskip 2mm}  
4274.119 & -5.330 & 0.490 & 1 & 9.5 & 0 & 10.5 & \multirow{3}{*}{76.2} \\
4274.133 & -3.025 & 0.074 & 0 & 6.5 & 0 &  6.5 & \\
4274.185 & -1.563 & 0.074 & 0 & 6.5 & 0 &  7.5 & \\
\noalign{\vskip 2mm}  
4356.284 & -5.228 & 0.389 & 1 & 4.5 & 0 & 5.5 & \multirow{6}{*}{68.4} \\
4356.343 & -3.529 & 0.343 & 1 & 1.5 & 0 & 0.5 & \\
4356.349 & -3.377 & 1.109 & 3 & 9.5 & 3 & 8.5 & \\
4356.355 & -1.846 & 0.157 & 0 & 8.5 & 0 & 7.5 & \\
4356.370 & -1.456 & 1.109 & 3 & 9.5 & 3 & 9.5 & \\
4356.389 & -3.312 & 0.157 & 0 & 8.5 & 0 & 8.5 & \\
\noalign{\vskip 2mm}  
4356.594 & -1.793 & 0.157 & 0 & 9.5 & 0 & 8.5 & 53.1  \\
\bottomrule
\end{tabularx}
\end{table}

\subsection{Optical A-X transitions} \label{sec:ResAX}
\subsubsection{Continuum normalisation}

The A-X transitions are located in a very busy part of the solar spectrum with an uncertain continuum position. This uncertainty has a significant impact on the derived carbon abundance from these transitions. The normalisation of the IAG center-to-limb observations is not perfect and the normalised intensities regularly exceed unity by a few percent, making it clear that the continuum has to be fit in combination with the line shapes. 

Therefore we have decided to determine a linear correction to the continuum of the IAG center-to-limb observations in the wavelength region of ±\SI{1.5}{\angstrom} around the theoretical line centers using our 3D-NLTE synthetic spectra. For this we model the surrounding atomic lines in LTE using line data extracted from VALD \citep{KupkaPiskunov1999, PiskunovKupka1995, RyabchikovaPiskunov2015}. Problematic spectral lines, which are not well reproduced by the models, are masked out via sigma-clipping, considerably improving the continuum fit. We allow the carbon abundance to be varied in this initial fitting phase, but we only keep the continuum parameters. Strictly speaking the continuum correction is model dependent, but we find that the continuum levels predicted by the \Must and \Stagger models are indistinguishable.
Having obtained the continuum correction we renormalise the continuum and determine the equivalent widths of the diagnostic lines by fitting a Gaussian line profile to the wavelength region of ±\SI{0.08}{\angstrom} around the theoretical line centers using \texttt{SciPy}'s \texttt{curve\_fit} function. We take great care to estimate realistic uncertainties on the resulting equivalent widths measurements. The dominant error in the equivalent widths measurements of all lines is the continuum normalisation. The details of this error analysis are discussed in \ref{sec:EWerror}.

\subsubsection{Abundance fitting and center-to-limb variation} \label{sec:opticalCLV}
\begin{figure*}
    \centering
    \includegraphics[width=\linewidth]{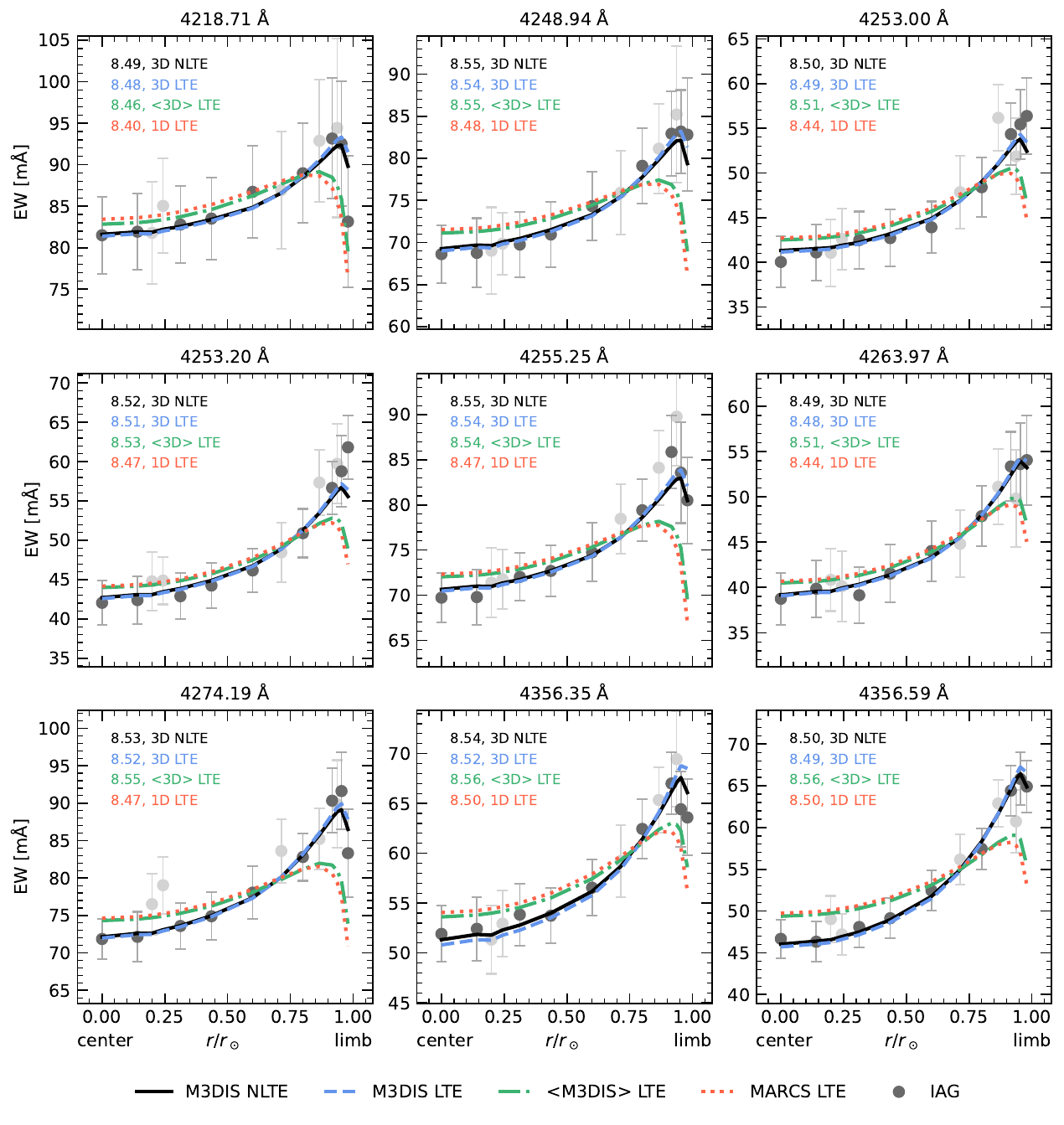}
    \caption{Synthetic equivalent widths of various models fitted to observed equivalent widths of the IAG center-to-limb observations. The x-axis shows relative distance from the disk center in solar radii. The dark grey dots represent the equivalent widths determined by Gaussian fits to the IAG spectra. The light grey dots are angles with very few observations, and hence low SNR, which have been omitted in the fitting procedure. Labels of the solid, dashed and dotted lines represent best-fit abundances for each model.}
    \label{fig:AX_CLV}
\end{figure*}

Having measured equivalent widths and their respective uncertainties for each IAG observation of a specific line, we fit the equivalent widths of the synthetic line profiles for all angles simultaneously. The employed line data for the A-X transitions is tabulated in Table \ref{tab:AX_data} and the resulting fits are presented in Fig. \ref{fig:AX_CLV}, which illustrates the results obtained for spatially resolved IAG solar spectra. For comparison, we also show the results obtained using the KPNO solar disc-center intensity \citep{brault1987spectral} and flux atlas \citep{KuruczFurenlid1984} in Appendix \ref{fig:CH_AX_KPNO}. The typical error on the abundance due to the fitting routine is $\pm0.04$ dex and it is approximately the same for all models as it is dominated by the uncertainty in the continuum placement. We note that overall the scatter of EW measurements is dominated by the properties of the observed IAG data (see also Fig. \ref{fig:IAG_CLV_4274} in Appendix). In particular, the errors bars are larger at angles closer to the limb, which is due to the reduced brightness at the limb. Also for selected angles closer to the centre, such as that at $\mu = 0.97, 0.98$ ($r/r_\odot=0.24, 0.20$), the observations are less reliable, because the data rely on a few observations only. This is, in particular, visible in the bottom left panel of Fig. \ref{fig:AX_CLV} that shows the 4274 \AA\ line. Clearly, the uncertainties of EW measurements are of the order $\sim 5$ percent, whereas the amplitude of the spatially-resolved data all CH lines spans a much larger range of over 20 percent and more importantly, \textit{the behaviour of the CH EWs is systematic, with EWs steadily increasing towards the limb}. Thus, the quality of the IAG data and the behaviour of the EWs as a function of angle is sufficient to enable detailed diagnostics of our line formation models for CH and of C abundances, as we demonstrate below. 

\begin{figure}
    \centering
    \includegraphics[width=\linewidth]{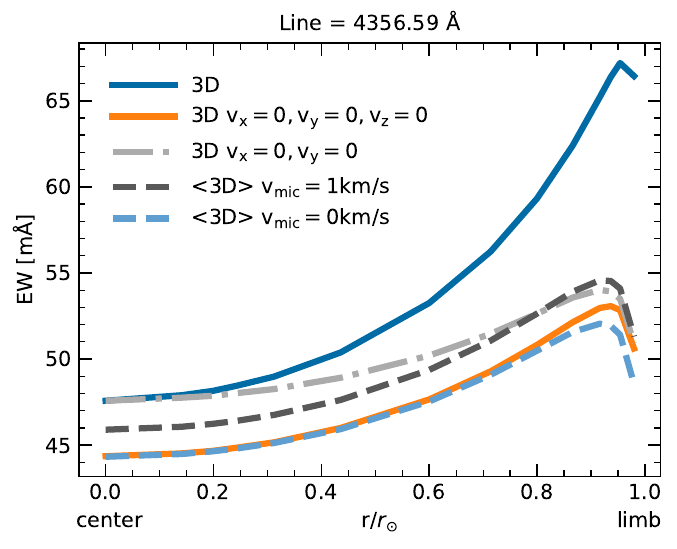}
    \caption{Equivalent widths as a function of distance to the disk center in solar radii. The solid blue curve shows the center-to-limb variation of a 3D hydrodynamical snapshot, which matches the solar observations well. The orange curve shows the result for the same snaphot and carbon abundance if all velocity components are manually set to zero. The grey dashed-dotted line shows the result of only setting the horizontal velocities to zero. The two dashed lines show the equivalent widths as produced by optically averaging the same snapshot and employing a constant microturbulence in the line formation calculations.}
    \label{fig:diffCLV}
\end{figure}

It is clearly visible that the CLV of the hydrostatic models, \Marcs and <\Must>, does not reproduce the observations nearly as well as the 3D \Must model in LTE and NLTE. The 3D hydrodynamic model predicts a steeper increase in the equivalent width towards the limb for all lines, however, the drop at the extreme limb ($r/r_\odot > 0.9$) is smaller than in the 1D models. This behaviour is examined more closely in Fig. \ref{fig:diffCLV}. There we show the CLV shapes for a single high-resolution 3D \Must snapshot and its optically averaged <3D> model, which is essentially a 1D model with the same average temperature and density stratification. All equivalent widths have been computed using the same carbon abundance. The present differences are purely due to the different treatments of velocities and temperature/density inhomogeneities. By comparison of the 3D model with all velocity components set to zero to the <3D> model without microturbulence we see that it is actually the hydrodynamical velocities, which are responsible for the considerably higher equivalent widths produced by the full 3D model across all angles. The temperature/density inhomogeneities appear to have little or no effect on the equivalent widths. Introducing microturbulence (arbitrary broadening of the opacity) increases the equivalent width in a systematic way, but fails to produce the steep increase towards the limb as seen in the hydrodynamical model. This increase can be attributed to the horizontal velocities in the snapshot, which cause significant broadening of the lines when viewing granulations cells side-on. The vertical velocities, on the other hand, account for an increase of the equivalent widths at the disk center.
\begin{figure}
    \centering
    \includegraphics[width=\linewidth]{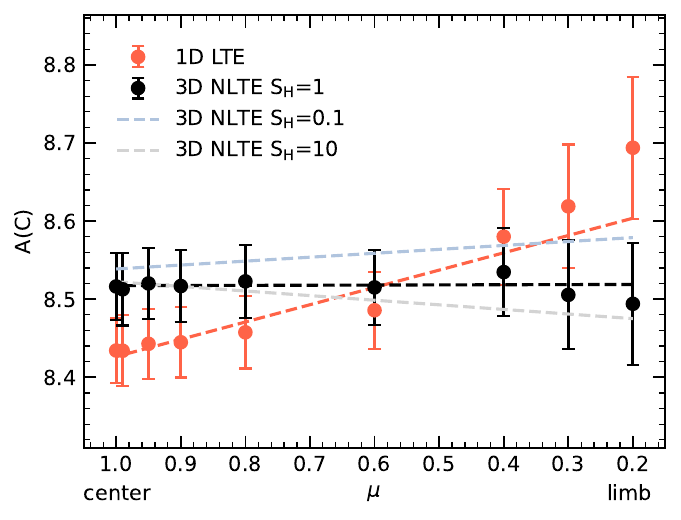}
    \caption{Derived carbon abundance as a function of observational $\mu$-angle. 1D LTE clearly results in different abundances for different $\mu$-angles, while the 3D NLTE result yields consistent results independently of the angle of observation. Scaling all hydrogen collision rates up or down by an order of magnitude breaks this consistency.}
    \label{fig:DrawinScaling}
\end{figure}
We have also tested how scaling the collision rates of CH$+$H affects the derived 3D NLTE carbon abundances. The combined result of all optical lines is shown in Fig. \ref{fig:DrawinScaling}. There is no abundance trend as a function of $\mu$-angle when the standard values are used. However, scaling all hydrogen collision rates up or down by an order of magnitude breaks this consistency.
In table \ref{tab:abundances} we combine our abundance estimates of the individual A-X transitions by taking their relative uncertainties into account. The result is a weighted mean, with the weights defined as $w=1/\sigma^2$. Our final carbon abundance estimate is $8.52\pm0.07$ dex when employing the \Must hydrodynamic atmosphere and dropping the assumption of LTE. In addition to the $0.04$ dex uncertainty associated with the fitting procedure, we incorporate a conservative uncertainty of $0.05$ dex due to the employed molecular and line data, namely molecular dissociation energies and log(gf) values of the diagnostic transitions. Fig. \ref{fig:DrawinScaling} reveals the uncertainty associated with the CLV $\sigma_{\rm CLV}$, which is $0.11$ dex in the 1D case and $0.06$ dex in the 3D case.
We consider the sources of error to be unrelated, hence we add them in quadrature and arrive at a total uncertainty of $\sigma_{\rm tot}=0.12$ dex for 1D and  $\sigma_{\rm tot}=0.07$ dex for 3D models. We show the synthetic line profiles for our final 3D NLTE and 1D LTE abundance estimates against the disk-center, limb and one intermediate observation in Fig. \ref{fig:CH_AX_fit} in the Appendix.
\begin{table*}
\setlength{\tabcolsep}{2pt}
\renewcommand{\footnoterule}{} 
\caption{Derived carbon abundances based on various model combinations. For lack of additional observation angles, we adopt $\sigma_{\rm CLV}$ from the optical also for the infrared transitions in the combined uncertainty $\sigma_{\rm tot}$.}
\label{tab:abundances}
\begin{center}
\begin{tabular*}{\linewidth}{@{\extracolsep{\fill}} p{2cm} c c | c c c | c c c c | c c c c}
\toprule
& & & \multicolumn{3}{c|}{3D Atmospheres} & \multicolumn{4}{c|}{1D Atmospheres} & \multicolumn{4}{c}{Uncertainty}   \\
  &   & & \Must & \Must & \Stagger & <\Must> & <\Cobold> & \Marcs & \Marcs   & $\sigma_{\rm fit}$ & $\sigma_{\rm CLV}$ & $\sigma_{\rm mol}$ & $\sigma_{\rm tot}$ \\
Lines & N$_\mu$ & N$_{\rm lines}$ & NLTE     & LTE      & LTE     & LTE        & LTE       & NLTE       & LTE & &  3D / 1D & & 3D / 1D \\
\midrule
A-X (optical) & 9 & 9 & 8.52 & 8.51 & 8.51 & 8.54 & 8.53 & 8.48 & 8.47 & 0.04 & 0.06 / 0.11 & 0.05 & 0.07 / 0.12 \\
vib-rot (infrared) & 1 & 50 & 8.51 & 8.51 & 8.51 & 8.52 & 8.51 & 8.49 & 8.48 & 0.01 & -\hspace{8pt}/\hspace{8pt}-  & 0.05 & 0.07 / 0.12 \\
\bottomrule        
\end{tabular*} 
\end{center}  
\end{table*}

\subsubsection{Non-LTE effects}
\begin{figure}
    \centering
    \includegraphics[width=\linewidth]{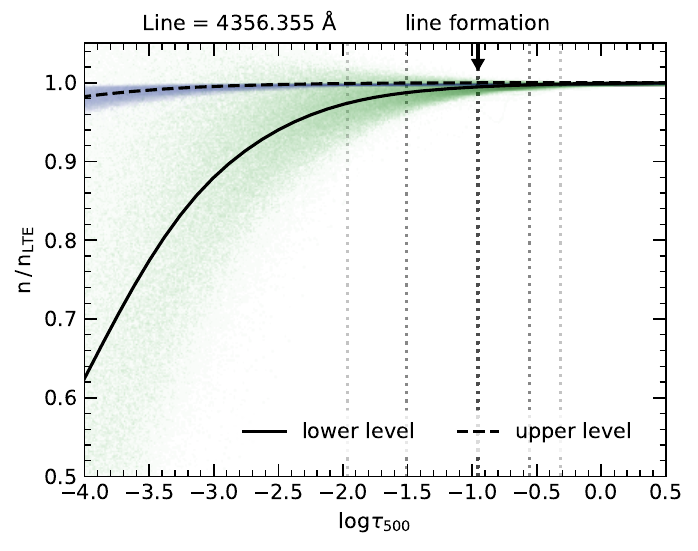}
    \includegraphics[width=\linewidth]{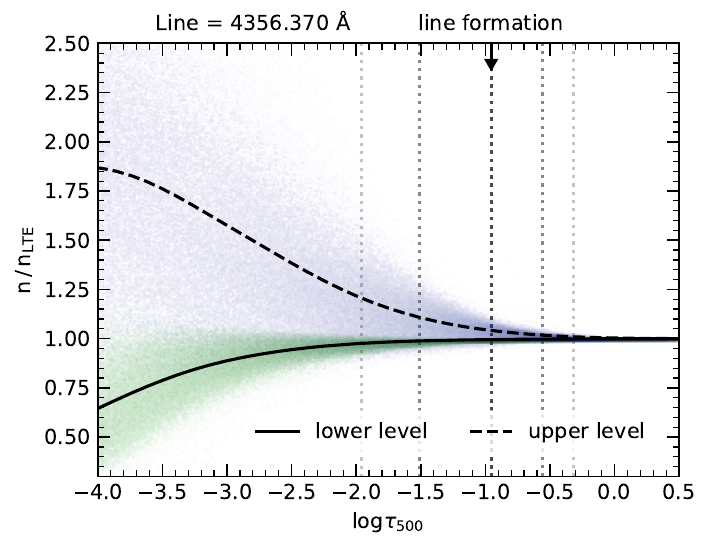}
    \caption{Departure coefficients of the lower and upper levels of the two strongest components of the \SI{4356.35}{\angstrom} line. The green shaded region represents the distribution of the lower level departure coefficients in the \Must simulation whereas the blue is for the upper level. Solid and dashed lines show the respective averages over surfaces of equal optical depths. We highlight the line forming regions by overplotting vertical dotted lines showing the optical depths where the contribution function at the line center reaches [33\%, 66\%, 100\%] of its maximum.}
    \label{fig:depart_4356}
\end{figure}

Under the LTE approximation we obtain a slightly lower abundance of $8.51$, which is the result of neglecting excitation and dissociation from photons. Fig. \ref{fig:depart_4356} shows the 3D departure coefficients (ratio of non-LTE to LTE level populations) of the two strongest components of the \SI{4356.35}{\angstrom} transition. The lower levels of the A-X transitions, which have low to intermediate excitation potentials with respect to the dissociation threshold, are generally underpopulated. The population of the upper energy levels depends on their energy relative to the dissociation threshold. It can be seen in Fig. \ref{fig:Grotrian} that some of the diagnostic A-X transitions cross the dissociation threshold. As levels get closer to the dissociation threshold their departure coefficients generally get closer to unity. Levels which lie well above the dissociation threshold, as is the case for the upper level of the \SI{4356.370}{\angstrom} transition, tend to become overpopulated. The result is in all cases an overestimation of the line strength in LTE, hence the lower LTE abundance. The overall effects are however small, as the CH lines form between $\log(\tau_{500})=-1$ and $-2$ (see Fig. \ref{fig:cntrbf2d} in the Appendix) where the departure coefficients are close to unity.

\subsection{Infrared rotation-vibration transitions}
We use a $\chi^2$-fit to match synthetic line profiles to the observations. The fitting procedure and error estimation is outlined in section \ref{sec:IRfitting} in the Appendix. The resulting 3D NLTE fits are presented in Fig. \ref{fig:IR_fits1} and \ref{fig:IR_fits2}. Abundance estimates for individual lines are also presented in the 3D NLTE and 1D LTE case in table \ref{tab:IR_data}.
We present the weighted mean abundances derived from each of the employed models in table \ref{tab:abundances}. The uncertainty on the combined abundance due to the fitting is estimated as $\sigma_{\rm fit} = (\sum 1/\sigma_i^2)^{-1/2}$.
For lack of data from additional observation angles, we adopt the same $\sigma_{\rm CLV}$ uncertainty as for the optical lines.
Similarly to the optical lines, we obtain a higher abundance of $8.51\pm0.07$ from the 3D NLTE fit compared to $8.48\pm0.12$ from the 1D LTE fit on average. The non-LTE effects on the rotation-vibration lines are below the $0.01$ dex level.

\section{Comparison with previous studies}
Our line selection overlaps with the one by AGS05 who focused on 9 A-X transitions and 102 infrared rotation-vibration transitions to determine the C abundance based on 3D LTE synthesis of CH lines. After the CH data by \cite{MasseronPlez2014} became available this analysis was updated for 7 out of the 9 A-X and 51 out of the 102 infrared transitions in \cite{AsplundAmarsi2021} (AAG21). Besides the updated line data AAG21 also employed the partition functions by \cite{BarklemCollet2016} and \Stagger snapshots with improved microphysics \citep{MagicCollet2013} compared to AGS05. Both studies supplied abundances based on a 1D \Marcs model for reference. An independent study by \cite{AlexeevaMashonkina2015} (AM15) also derived the solar carbon abundance employing \Marcs models and investigating the same 9 A-X transitions as AGS05.

\subsection{Optical A-X transitions}
AGS05 derived a 3D LTE abundance of $8.45\pm0.04$ based on the KPNO solar disk-center atlas ($\mu=1.0$). This value was slightly revised to $8.46\pm0.05$ in AAG21, reflecting the use of updated atmospheric and molecular data and the omission of two of the nine A–X transitions. The specific transitions omitted were not identified in their study. While the change in the 3D abundance is small, the corresponding 1D LTE abundance decreased from $8.44\pm0.04$ to $8.39\pm0.05$ between the two analyses. The different behaviour of the 1D and 3D results is somewhat surprising, given the substantial updates to the molecular data, and may point to a compensating effect arising from the change in the underlying 3D \Stagger model between AGS05 and AAG21. However, we were not able to find a detailed discussion in Asplund et al. of how the updates to their 3D solar model affect the formation of the A–X CH lines. This suggests that the updated \Stagger atmosphere could have introduced a significant upward correction of the carbon abundance for the A–X transitions. A direct differential comparison between our 3D LTE result and that of AAG21 is unfortunately not possible, as the necessary 3D atmosphere models are not publicly available and the set of seven lines retained in their analysis was not specified. In addition, it remains unclear which of the observational data mentioned in AAG21 were used in their optical-line analysis. Since the choice of observations (disk-center intensity, flux, or specific angular positions) can influence the derived carbon abundance (as illustrated in Fig. \ref{fig:CH_AX_KPNO} in the Appendix) this information would be valuable for a more complete comparison.

The results of the 1D LTE fits presented in Fig. \ref{fig:CH_AX_KPNO} are within the uncertainty of the AAG21 1D LTE result, except for the \SI{4356.35}{\angstrom} and the \SI{4356.59}{\angstrom} lines, which do not have to be included in their line sample. However, this difference is easily accounted for by a continuum renormalisation, as the continuum placement is rather uncertain due to the many blending lines. As the CLV analysis accounts for uncertainties in the continuum placement and does not produce outlying abundances for the two lines, we deem the source of this issue unlikely to be an unidentified background line.
Employing the KPNO solar flux atlas AM15 arrive at $8.39\pm0.02$ in their 1D LTE analysis. Using the same flux atlas \ref{fig:CH_AX_KPNO} (bottom panel), we find mostly larger abundances by $\sim0.03$ dex (irrespective of the \SI{4356}{\angstrom} doublet). Neither AAG21 nor AM15 published their final line profiles and best-fit models. But one may suggest that the remaining differences are due to differences in the broadening recipes, background linelist, continuum renormalisation or the chosen fitting windows.

\subsection{Infrared rotation-vibration transitions}
\label{sec:IRlineSelection}
AAG21 state a 3D LTE abundance of $8.47\pm0.04$ as derived from 51 of the 102 rotation-vibration transitions employed in AGS05. Unfortunately there is no mention of how the original sample from AGS05 was cut in half. As the line-to-line scatter in AGS05 covers a range of roughly $0.2$ dex (Fig 4. in \citep{AsplundGrevesse2005}) a systematic reduction of the sample can easily account for the difference between their result and our 3D LTE abundance of $8.51\pm0.07$ derived with a solar \Stagger simulation similar to theirs. Additionally, the continuum level in the raw Spacelab-3 \texttt{ATMOS} data frequently exceeds unity (see Fig. \ref{fig:IR_fits3} in the Appendix), which contributes another source of significant uncertainty as custom renormalisation is inevitable.

\subsection{Granulation effects on CH number densities} \label{sec:molecules}
As shown in Fig \ref{fig:diffCLV} we find only small differences in the EWs of the CH lines between the full 3D \Must model and its optically averaged <3D> version when removing all velocity components from the line formation calculation. This means that inhomogeneities in the full 3D model have a negligible effect on the CH line strength, at least in the Sun. This is an initially surprising result as \cite{UitenbroekCriscuoli2011} conclude that <3D> models always underestimate molecular number densities present in stellar atmospheres. They showcase this employing the CO molecule and find confirmation in the studies by \cite{KiselmanNordlund1995} and \cite{ScottAsplund2006}, who derive larger abundances from <3D> models compared to full 3D models for OH and CO lines, respectively.

To explore this interesting behaviour further, we compare the difference between the <3D> number densities $N_{\rm <3D>}$ and the full 3D number densities averaged over surfaces of equal optical depth <$N_{\rm 3D}$> for different molecules using a single \Must 3D RHD snapshot (see Fig. \ref{fig:molecules} in the Appendix). Using the <3D> (time and spatially averaged) atmosphere results in underestimating the number densities in the line forming regions for the CO, OH, NH and O$_2$ molecules. However, in the case of CH and C$_2$ the number densities are overestimated at $\log\tau < -0.5$. The reason for this is the high dissociation energy of the CO molecule, which is roughly three times as large as the dissociation energy of CH. This causes a substantial fraction of the total number of C atoms $N_{\rm C}$ to be locked in CO in this optical depth range. Below a certain temperature, the fraction of C locked in CH no longer increases with decreasing temperature, but is reduced due to $N_{\rm CO}/N_{\rm C}$ approaching unity. Thus, the number density of C\,\textsc{i} and therefore CH becomes anti-correlated to the number density of CO and the approximation -- atomic number densities are constant under small perturbations in temperatures -- used in \cite{UitenbroekCriscuoli2011} breaks down.

\subsection{Solar Carbon abundance}
We adopt the C abundance of $8.52\pm0.07$ as our final estimate of the solar C abundance. This abundance represents the weighted average of the two results from our optical and infra-red analyses. The analysis of the optical CH A-X transitions was based on 81 observations in total, 9 observations for each of the 9 lines, and the analysis of the 50 infra-red vibration-rotation transitions was based on a single observation for each line.

The most recent analysis of atomic C\,\textsc{i} lines in the solar photosphere was conducted by \cite{AmarsiBarklem2019} (hereafter AB19), using 3D NLTE radiative transfer models similar to those employed in this work. AB19 determined a C abundance of $8.44\pm0.02$, which is notably lower than our measurement of $8.52\pm0.07$, despite their small uncertainties. However, AB19 noted that adopting the line selection and equivalent width measurements from \cite{CaffauLudwig2010} would increase their derived C abundance by $0.07$ dex. \cite{CaffauLudwig2010} performed a 3D LTE analysis of the solar C\,\textsc{i} diagnostics using \Cobold models. Furthermore, updated theoretical calculations of transition data for C ions by \cite{LiAmarsi2021} yielded a C abundance of $8.50\pm0.07$ when the oscillator strengths from their study were applied to the AB19 analysis. AB19 state that 1D to 3D abundance differences (3D typically yielding higher abundances) are caused by differences in the atmospheric stratification finding that 3D-⟨3D⟩ abundance corrections are much less severe (<0.03 dex at disk-centre). \cite{MaggBergemann2022} find a C abundance of $8.56\pm0.05$ for C\,\textsc{i} lines assuming LTE and employing an <3D> \Stagger model. This result agrees with our findings, especially when applying the 3D NLTE - 3D LTE abundance difference of $-0.01$ dex found by AB19. 
In sum, we show that our 3D NLTE C abundance of $8.52\pm0.07$ based on molecular CH lines is in agreement with the C\,\textsc{i} measurements.

\section{Discussion}
\begin{figure}
    \centering
    \includegraphics[width=\linewidth]{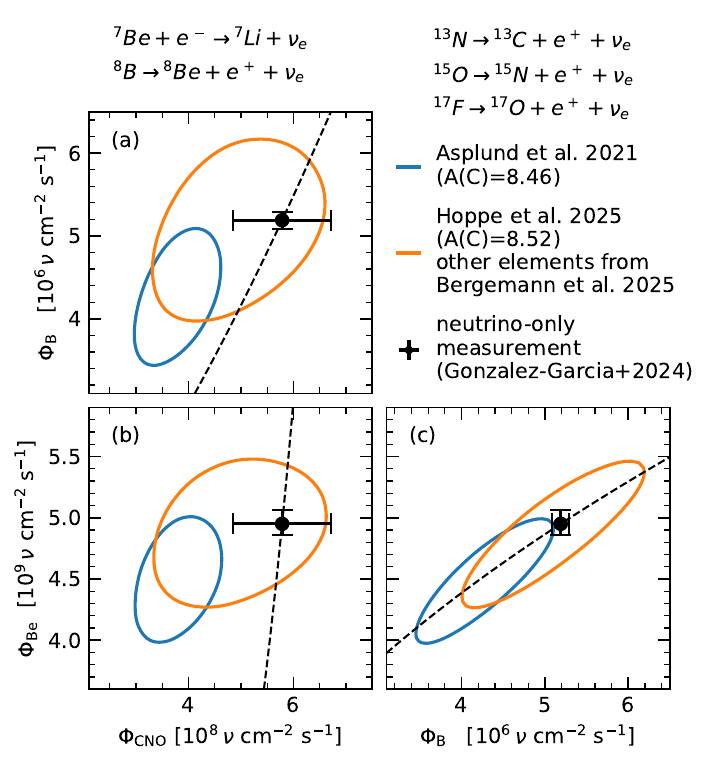}
    \caption{Neutrino fluxes from global analysis of experimental neutrino measurements (black circles). The contours show predicted neutrino fluxes based on standard solar models with varying solar compositions. The black dashed tracks show the correlation between the fluxes and the solar core temperature.}
    \label{fig:neutrino_fluxes}
\end{figure}

Our new 3D NLTE solar C abundance is highly relevant within the context of Standard Solar Models (SSMs). These models are used, among other applications, to predict the rates of nuclear fusion processes in the Sun's core \citep[e.g.][]{VinyolesSerenelli2017}. The Sun generates energy through two primary mechanisms: the proton-proton (pp) chain, which is responsible for approximately 99\% of the energy, and the much subdominant carbon-nitrogen-oxygen (CNO) cycle. Both processes produce neutrinos that easily escape the core, providing a direct probe of the physics of the Sun's interior. Importantly, the CNO cycle, catalyzed by the abundances of C and N in the core, yields distinct neutrino fluxes (specifically from the decay of $^{13}$N, $^{15}$O, and $^{17}$F, Fig. \ref{fig:neutrino_fluxes}) that are nearly linearly dependent on the sum of abundances of C and N in the core \citep[e.g.][]{Serenelli2013, AppelBagdasarian2022, BasilicoBellini2023}. These neutrino fluxes, along with those produced in the pp chain can be probed in experiments, such as Borexino \citep{BorexinoCollaborationAltenmuller2020, AppelBagdasarian2022}. To explore how the properties of the solar core change with our new 3D NLTE A(C), we use our well-established and carefully-tested SSM \citep{Bahcall2006, SerenelliBasu2009, SerenelliHaxton2011, VinyolesSerenelli2017} to compute theoretical neutrino fluxes for the Sun. For other elements, we use our data from \citet{BergemannLodders2025} and \citet{LoddersBergemann2025} (hereafter, BLP25).

Fig. \ref{fig:neutrino_fluxes} shows our predicted neutrino fluxes for $^{7}$Be, $^{8}$B, and for the total CNO cycle ($^{13}$N+$^{15}$O+$^{17}$F). The plotted contours outline the 68\% confidence levels of our SSM-based neutrino fluxes. Here we also overplot the  experimental data from the global analysis by \cite{Gonzalez-GarciaMaltoni2024}. As suggested by the authors, we use their results derived without imposing the solar luminosity constraint\footnote{We note that imposing the solar luminosity constraint leads to an overall reduction of neutrino fluxes from the Sun. However, it effectively leads to an increased carbon abundance, due to an anti-correlation between the fluxes released by the proton-electron-proton (pep) reaction and by the CNO cycle.}, as these results provide a purely measurement-based reference and do not rely on the SSM. We also include, for visual comparison, our results for the neutrino fluxes obtained using the composition by \citet{AsplundAmarsi2021}. As seen in this Fig., the experimental solar neutrino fluxes are consistent with the predictions of the SSM based on our solar 3D NLTE A(C) and BLP25 abundances. However, there is a systematic discrepancy with the SSM neutrino results based on AAG21 composition, which has also been demonstrated and extensively discussed in \citet{AppelBagdasarian2022} and \citet{BasilicoBellini2023}. We note that BLP25 provides more conservative errors on individual abundance measurements, hence the contours cover the largest areas.

We note that the neutrino fluxes produced in a given nuclear reaction are strongly correlated with the temperature in the solar core $T_c$, which is an output of the SSM, and with the abundances of the elements \citep{BahcallUlmer1996}. The black dashed lines in Fig. \ref{fig:neutrino_fluxes} depict how the fluxes change with the change of $T_c$. This is crucial, because the uncertainties of all other input parameters of the SSM including abundances, nuclear reaction rates, and environmental parameters ($L_\odot$, opacity, age, and diffusion) enter the predicted fluxes from $^{7}$Be and $^{8}$B \textit{only through $T_c$}. The only exception being the reactions $^{7}$Be+p and $^{7}$Be+e. However, these nuclear rates are well-understood and the measurement uncertainties are within $5\%$ \citep{AcharyaAliotta2025}. Hence, the impressive agreement between our model fluxes and the experimental data for $^{7}$Be and $^{8}$B fluxes  - to better than 10\% - indicates that the solar $T_c$ is well reproduced. In the case of CNO fluxes, once $T_c$ is fixed by the good agreement of $^{7}$Be and $^{8}$B, only two remaining factors can reconcile the AAG21-based SSM model predictions with the CNO measurement: (i) a higher C+N abundance, or (ii) a change in the $^{14}$N+p reaction rate. However, since the CNO neutrino flux depends almost linearly on this rate, a  $\sim40\%$ increase of the rate would be necessary to match the observations. This large difference is well-beyond the uncertainty of this reaction rate of $\sim8\%$ \citep{AcharyaAliotta2025}. For a detailed analysis of the SSM we refer the reader to our earlier papers \citep[e.g.][and references therein]{Serenelli2013, VinyolesSerenelli2017}.

\begin{figure}
    \centering
    \includegraphics[width=\linewidth]{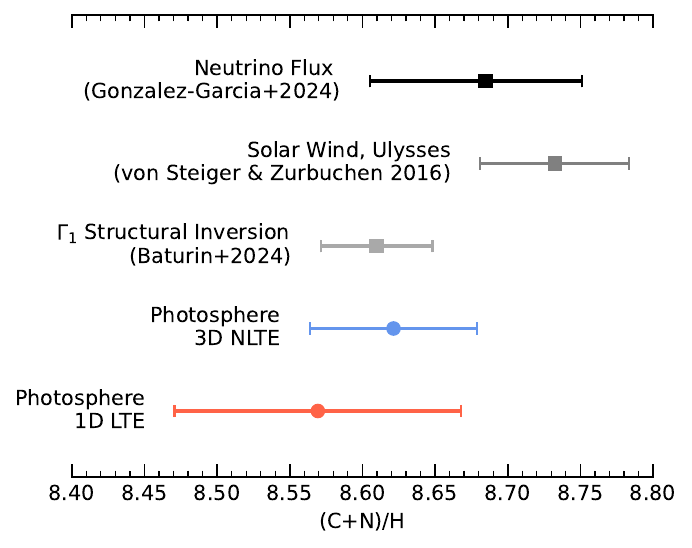}
    \caption{Combined carbon and nitrogen abundance measurements from different components of the Sun. We adopted the nitrogen abundances of $7.94\pm0.11$ \citep[3D NLTE,][]{BergemannLodders2025} and $7.88\pm0.12$ \citep[1D LTE,][]{MaggBergemann2022} as the respective photospheric values. Here the 'Neutrino flux' data  correspond to the independent solar surface C+N value, obtained by combining the neutrino fluxes measured by the Borexino experiment \citep{Gonzalez-GarciaMaltoni2024} and the SSM model \citep{Serenelli2013}.}
    \label{fig:Borexino}
\end{figure}

Finally, in Fig. \ref{fig:Borexino} we compare spectroscopic 3D NLTE result for A(C)  with other estimates. To illustrate the difference of using 3D NLTE instead of 1D LTE, we also show our A(C) obtained from 1D LTE modelling of the solar CH lines. Here we adopt the C/N ratio from BLP25, A(C+N)=$8.68_{-0.08}^{+0.07}$. For comparison, we show the \textit{photospheric} C+N value based on combining the CNO neutrino fluxes from \citet{Gonzalez-GarciaMaltoni2024} with the SSM, as described above. We also show the results obtained from the analysis of the solar wind data from polar coronal holes collected by the ESA's \texttt{Ulysses} solar space probe \citep{vonSteigerZurbuchen2016}. These yield the C abundance of $8.65\pm0.06$ \citep{vonSteigerZurbuchen2016}. Finally, we include the recent estimate of C and N obtained from modelling the adiabatic $\Gamma_1$ index of the solar interior \citep{BaturinOreshina2024}. They find A(C)=$8.44\pm0.04$ and A(N)=$8.12\pm0.08$. Also these measurements are consistent with our estimate within the combined uncertainties of both values.


\section{Conclusions}
In this work, we explore the influence of 3D NLTE effects on the formation of diagnostic absorption lines of the CH molecule  in the solar spectrum. Our linelist includes the optical CH lines at \SI{4218}{}-\SI{4356}{\angstrom} and the far-IR CH lines at \SI{33025}{}-\SI{37944}{\angstrom}). Our goal is to test the influence of these effects on the analysis of solar photospheric C abundance. This quantity is particularly relevant in studies of the structure of the solar interior \citep{SerenelliBasu2009, MaggBergemann2022} and in the context of neutrino measurements \citep{AppelBagdasarian2022, BasilicoBellini2023}.

We model NLTE radiation transfer in the CH molecule using 3D radiation-hydrodynamics models of the solar atmosphere from our previous work \citep{EitnerBergemann2024} and from the \Stagger grid \citep{MagicCollet2013}. We employ an updated version of the CH model molecule from \cite{PopaHoppe2023}, but grouping individual energy states into superlevels. We have verified carefully that our level averaging procedure does not compromise NLTE effects. For a comparative analysis, we calculate 1D LTE, 1D NLTE, <3D> LTE, 3D LTE and full 3D NLTE CH line profiles. The resulting synthetic spectra are compared with the observed high-resolution solar flux data as well as with the spatially-resolved solar spectra IAG taken at multiple angles across the solar disc \citep{EllwarthSchafer2023}.

We find that 1D and average 3D LTE models are unable to describe the observed centre-to-limb variation, leading to underestimated (by -0.1 dex) C abundance at the disk centre and highly overestimated (by over 0.2 dex) C abundances at the solar limb. In contrast, 3D LTE and 3D NLTE models yield consistent abundances at all pointings across the solar disc. Furthermore, 3D models improve the agreement with the observed line profiles, compared to 1D. Our 3D NLTE estimate is A(C)$_{\rm 3D\,NLTE} = 8.52\pm0.07$ based on the weighted average of optical and infra-red CH lines. In comparison, we obtain A(C)$_{\rm 3D\,LTE} = 8.51\pm0.07$, A(C)$_{\rm MARCS\,NLTE} = 8.49\pm 0.12$,  A(C)$_{\rm MARCS\,LTE} = 8.49\pm 0.12$. The error reflects uncertainties of the data-model comparison, the systematics with $\mu$, and that of the employed atomic and molecular data (oscillator strengths, partition functions, and CH dissociation energy). The choice of 3D radiation-hydrodynamics atmosphere model does not impact our result. We obtain identical C abundances using the 3D RHD \Must and \Stagger solar models.

By comparing the line formation in 3D LTE and 3D NLTE, we find that NLTE effects on the CH optical and far-IR lines are very small and amount to only $\sim 0.01$ dex (Table \ref{tab:abundances}). The NLTE effects in the optical and IR diagnostic lines are caused by underpopulation of the lower energy states. Our analysis of the impact of the rather uncertain $C+H$ collision rates \cite[][]{Drawin1969} on abundances confirms that the impact is very small. Using the scaling factor of S$_H =0.1$ and S$_H =10$ in statistical equilibrium calculations leads to A(C) higher by $0.03$ and lower by $0.01$ dex, respectively. However, with these scaling factors, the abundances become more discrepant as a function of the $\mu$ angle. We thus favour the unscaled collisional rates in the model.

Our 3D NLTE estimate based on CH lines is in good agreement with the estimate based on atomic C\,\textsc{i} lines from \cite{MaggBergemann2022}, which is 8.56$\pm$0.05 dex. Interestingly, our estimate is also consistent with the surface C abundance ($8.58_{-0.08}^{+0.07}$) inferred from the global analysis of the solar neutrino flux measurements \citep{Gonzalez-GarciaMaltoni2024}. Additionally, our 3D NLTE value aligns more closely with the value of $8.65 \pm 0.07$, derived from solar wind measurements based on Ulysses data \citep{vonSteigerZurbuchen2016}, than the 1D LTE result.

With this paper, we release the model molecule, our 3D Dispatch model atmosphere is public \citep[][\footnote{\href{https://nlte.mpia.de}{\nolinkurl{https://nlte.mpia.de}}}]{EitnerBergemann2024} and the line list is provided in tables \ref{tab:AX_data} and \ref{tab:IR_data}. Finally we are preparing a dedicated paper to release our updated 3D NLTE spectrum synthesis code.

\section*{Acknowledgements}
M.B. is supported through the Lise Meitner grant from the Max Planck Society. This project has received funding from the European Research Council (ERC) under the European Unions Horizon 2020 research and innovation programme (Grant
agreement No. 949173). B.P. is supported in part by the national French space agency (CNES). A.S. acknowledges support by the Spanish Ministry of Science, Innovation and Universities through the grant PID2023-149918NB-I00 and the program Unidad de Excelencia Mar\'{i}a de Maeztu CEX2020-001058-M, and Generalitat de Catalunya through grant 2021-SGR-1526, and the ChETEC-INFRA H2020 project under contract No 101008324. Finally, we thank Luc Rouppe Van Der Voort for locating and retrieving observations from the Swedish 1-m Solar Telescope (SST) archive. The SST is operated on the island of La Palma by the institute for Solar Physics of Stockholm University in the Spanish Observatorio del Roque de los Muchachos of the Instituto de Astrofisica de Canarias. The SST is co-funded by the Swedish Research Council as a national research infrastructure (registration number 4.3-2021-00169).

\section*{Data Availability}
The Spacelab-3 \texttt{ATMOS} atlas ($\mu=0.935$) \citep{FarmerNorton1989} as uploaded by \cite{SeoKim2007} (file \texttt{solspec2b\_1.txt}) but renormalised using \texttt{SUPPNet} \citep{RozanskiNiemczura2022} will be made available in machine-readable format via CDS upon publication of the paper. The employed model of the CH molecule is uploaded as supplementary material to the paper as well.



\bibliographystyle{mnras}
\bibliography{CH_Sun} 




\appendix

\section{}

\begin{table}
\centering
\caption{Infrared vibration-rotation line data and resulting 3D NLTE (\Must) and 1D LTE (\Marcs) abundances from profile fitting procedure. $E_i$ represents the lower excitation potential of each transition. $W_{\rm eq}$ represents the equivalent widths measured for each line in the Spacelab-3 \texttt{ATMOS} atlas.}
\label{tab:IR_data}
\begin{tabularx}{\linewidth}{
>{\centering\arraybackslash}X@{}
>{\centering\arraybackslash}X@{}
>{\centering\arraybackslash}X@{}
>{\centering\arraybackslash}X@{}
>{\centering\arraybackslash}X
>{\centering\arraybackslash}X}
\toprule
$\lambda$ [\AA] & $\log(g\!f)$ & $E_i$ [eV] & $W_{\rm eq}$ [m\AA] & 3D NLTE & 1D LTE  \\
\midrule
33025.10 & -3.102 & 0.465 & 12.27 & 8.51 $\pm$ 0.04 & 8.46 $\pm$ 0.05 \\
33025.92 & -3.063 & 0.465 & 13.03 & 8.49 $\pm$ 0.04 & 8.46 $\pm$ 0.05 \\
33272.73 & -3.234 & 0.315 & 12.11 & 8.50 $\pm$ 0.03 & 8.48 $\pm$ 0.05 \\
33274.10 & -3.189 & 0.315 & 13.06 & 8.50 $\pm$ 0.03 & 8.47 $\pm$ 0.04 \\
33398.58 & -3.283 & 0.270 & 13.97 & 8.58 $\pm$ 0.03 & 8.55 $\pm$ 0.04 \\
33534.27 & -3.335 & 0.229 & 12.01 & 8.52 $\pm$ 0.03 & 8.49 $\pm$ 0.05 \\
33535.90 & -3.285 & 0.229 & 12.96 & 8.51 $\pm$ 0.03 & 8.48 $\pm$ 0.04 \\
33850.47 & -3.450 & 0.157 & 11.12 & 8.53 $\pm$ 0.04 & 8.50 $\pm$ 0.05 \\
33852.97 & -3.393 & 0.157 & 11.73 & 8.51 $\pm$ 0.03 & 8.47 $\pm$ 0.05 \\
33859.39 & -3.393 & 0.157 & 12.90 & 8.53 $\pm$ 0.03 & 8.51 $\pm$ 0.04 \\
34038.40 & -3.516 & 0.126 & 10.25 & 8.53 $\pm$ 0.04 & 8.50 $\pm$ 0.05 \\
34041.38 & -3.453 & 0.126 & 10.41 & 8.48 $\pm$ 0.04 & 8.45 $\pm$ 0.05 \\
34044.61 & -3.516 & 0.125 & 11.20 & 8.55 $\pm$ 0.03 & 8.54 $\pm$ 0.05 \\
34047.42 & -3.453 & 0.126 & 11.61 & 8.51 $\pm$ 0.03 & 8.50 $\pm$ 0.05 \\
34244.07 & -3.588 & 0.098 &  8.46 & 8.50 $\pm$ 0.04 & 8.47 $\pm$ 0.06 \\
34247.76 & -3.518 & 0.098 & 10.24 & 8.51 $\pm$ 0.04 & 8.48 $\pm$ 0.05 \\
34249.82 & -3.588 & 0.098 &  8.80 & 8.51 $\pm$ 0.04 & 8.48 $\pm$ 0.06 \\
34253.37 & -3.518 & 0.098 & 10.62 & 8.51 $\pm$ 0.04 & 8.50 $\pm$ 0.06 \\
34477.50 & -3.591 & 0.074 &  9.23 & 8.51 $\pm$ 0.04 & 8.48 $\pm$ 0.06 \\
34802.64 & -2.890 & 0.686 & 11.82 & 8.52 $\pm$ 0.03 & 8.49 $\pm$ 0.05 \\
34803.84 & -2.848 & 0.686 & 13.24 & 8.53 $\pm$ 0.03 & 8.50 $\pm$ 0.04 \\
34896.24 & -2.934 & 0.642 & 11.07 & 8.50 $\pm$ 0.04 & 8.47 $\pm$ 0.05 \\
34897.71 & -2.891 & 0.642 & 11.33 & 8.49 $\pm$ 0.03 & 8.44 $\pm$ 0.05 \\
35017.46 & -2.936 & 0.599 & 12.75 & 8.52 $\pm$ 0.03 & 8.49 $\pm$ 0.05 \\
35023.20 & -2.982 & 0.598 & 10.33 & 8.49 $\pm$ 0.04 & 8.45 $\pm$ 0.05 \\
35024.79 & -2.936 & 0.599 & 11.78 & 8.50 $\pm$ 0.03 & 8.46 $\pm$ 0.04 \\
35153.74 & -3.032 & 0.560 & 10.62 & 8.50 $\pm$ 0.04 & 8.47 $\pm$ 0.05 \\
35155.72 & -2.984 & 0.560 & 11.81 & 8.50 $\pm$ 0.03 & 8.47 $\pm$ 0.05 \\
35162.95 & -2.984 & 0.559 & 12.24 & 8.50 $\pm$ 0.03 & 8.48 $\pm$ 0.05 \\
35485.73 & -3.145 & 0.490 &  9.53 & 8.50 $\pm$ 0.04 & 8.47 $\pm$ 0.06 \\
35488.52 & -3.089 & 0.490 & 11.28 & 8.51 $\pm$ 0.04 & 8.48 $\pm$ 0.05 \\
35680.18 & -3.210 & 0.460 &  9.13 & 8.50 $\pm$ 0.05 & 8.48 $\pm$ 0.06 \\
35683.44 & -3.148 & 0.460 & 10.14 & 8.51 $\pm$ 0.04 & 8.47 $\pm$ 0.05 \\
35899.90 & -3.281 & 0.433 &  8.34 & 8.50 $\pm$ 0.05 & 8.49 $\pm$ 0.07 \\
36491.71 & -2.649 & 1.043 &  9.50 & 8.52 $\pm$ 0.04 & 8.50 $\pm$ 0.06 \\
36498.21 & -2.687 & 1.042 &  7.90 & 8.50 $\pm$ 0.05 & 8.46 $\pm$ 0.07 \\
36499.41 & -2.649 & 1.043 &  8.43 & 8.49 $\pm$ 0.05 & 8.46 $\pm$ 0.07 \\
36655.99 & -3.568 & 0.372 &  5.54 & 8.53 $\pm$ 0.08 & 8.53 $\pm$ 0.13 \\
36678.14 & -2.729 & 0.952 &  8.93 & 8.51 $\pm$ 0.04 & 8.48 $\pm$ 0.07 \\
36927.54 & -2.864 & 0.874 &  7.98 & 8.52 $\pm$ 0.05 & 8.49 $\pm$ 0.08 \\
36929.66 & -2.817 & 0.874 &  8.61 & 8.50 $\pm$ 0.05 & 8.47 $\pm$ 0.07 \\
36935.24 & -2.864 & 0.874 &  8.29 & 8.52 $\pm$ 0.05 & 8.50 $\pm$ 0.08 \\
36937.18 & -2.817 & 0.874 &  9.23 & 8.52 $\pm$ 0.05 & 8.50 $\pm$ 0.07 \\
37087.34 & -2.916 & 0.839 &  7.45 & 8.51 $\pm$ 0.06 & 8.47 $\pm$ 0.08 \\
37094.82 & -2.916 & 0.839 &  7.47 & 8.52 $\pm$ 0.06 & 8.47 $\pm$ 0.08 \\
37097.14 & -2.865 & 0.839 &  8.26 & 8.50 $\pm$ 0.05 & 8.47 $\pm$ 0.07 \\
37690.07 & -3.107 & 0.753 &  6.47 & 8.53 $\pm$ 0.06 & 8.52 $\pm$ 0.09 \\
37700.78 & -3.038 & 0.753 &  6.51 & 8.51 $\pm$ 0.05 & 8.46 $\pm$ 0.07 \\
37932.89 & -3.187 & 0.730 &  5.99 & 8.59 $\pm$ 0.07 & 8.54 $\pm$ 0.09 \\
37944.25 & -3.109 & 0.730 &  7.08 & 8.54 $\pm$ 0.07 & 8.53 $\pm$ 0.11 \\
\bottomrule
\end{tabularx}
\end{table}

\begin{figure}
    \centering
    \includegraphics[width=\linewidth]{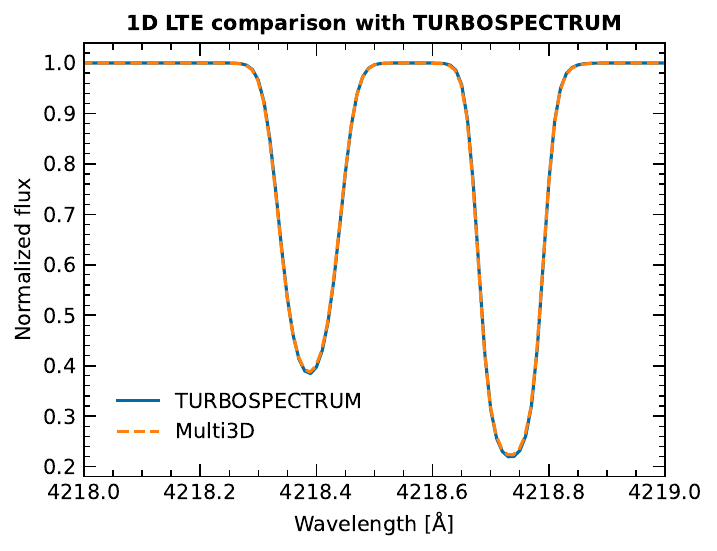}
    \caption{Comparison of synthetic CH line profiles using our custom version of \MultiD and the independent spectrum synthesis code \Turb. This comparison has been done in 1D LTE, as \Turb is a 1D LTE code.}
    \label{fig:TScomparison}
\end{figure}

\begin{figure}
    \centering
    \includegraphics[width=\linewidth]{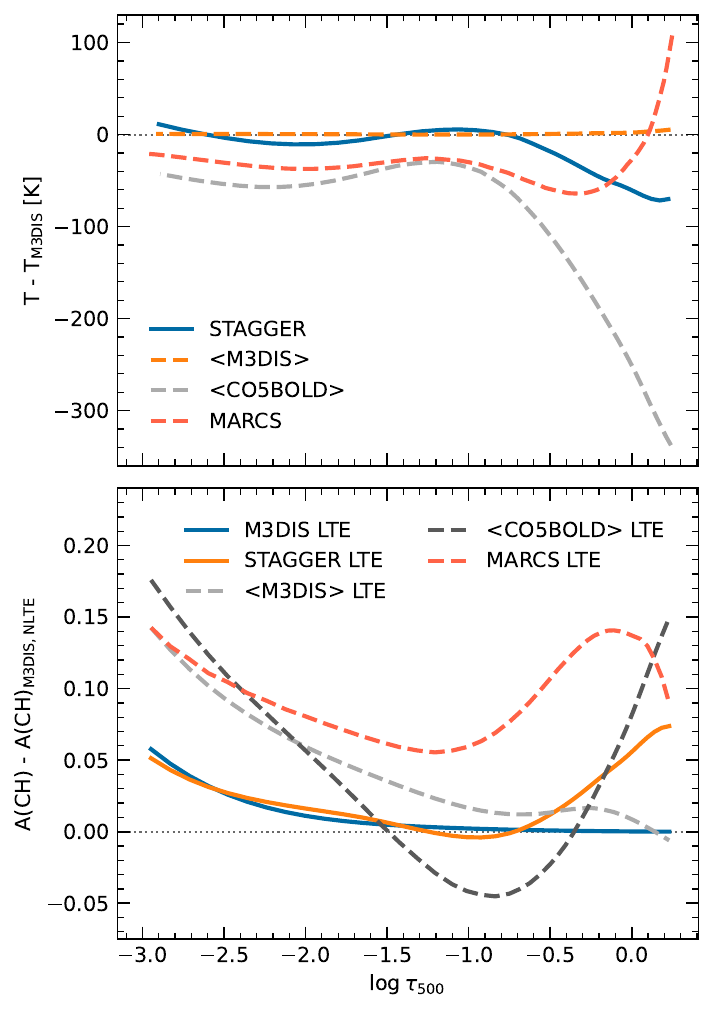}
    \caption{Top panel: temperature structure comparison of utilised atmosphere models in the critical line forming region around $\log \tau_{500}=-1$. Bottom panel: Resulting CH abundances, defined as $\rm \log(N_{CH}/N_{H})+12$, as a function of optical depth.}
    \label{fig:atmos_comparison}
\end{figure}

\begin{figure}
    \centering
    \includegraphics[width=\linewidth]{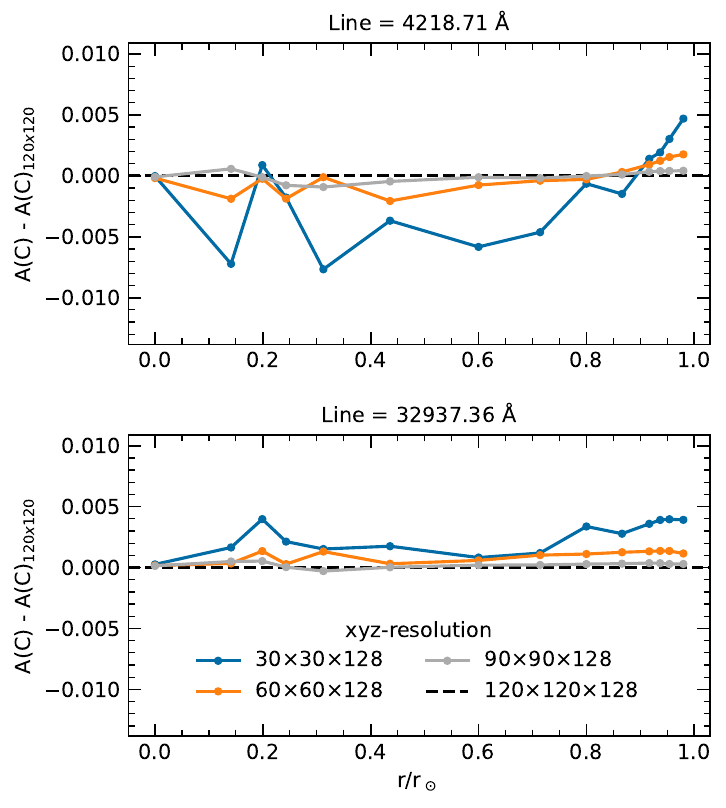}
    \caption{Impact of resolution of interpolated snapshots on the determined abundances for one optical (top panel) and one infrared (bottom panel) line. The effect is minimal for the vertical ray at the disk center (r=0), but also maximal for rays which are nearly vertical. The full snapshot resolution is $240\times240\times239$.}
    \label{fig:resolution}
\end{figure}

\begin{figure}
    \centering
    \includegraphics[width=\linewidth]{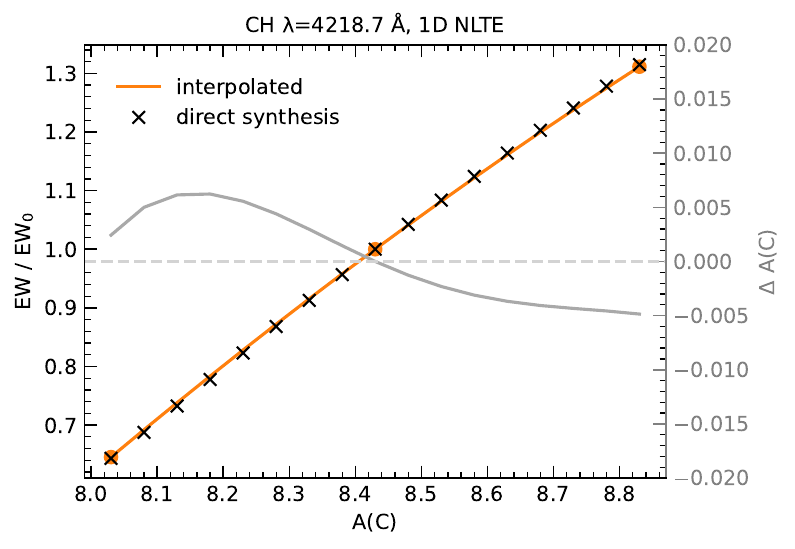}
    \includegraphics[width=\linewidth]{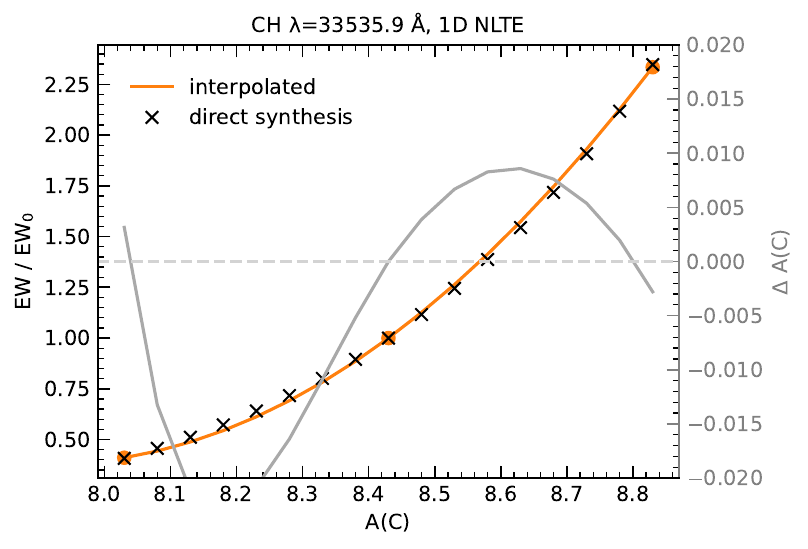}
    \caption{The two panels show the curve of growth of the strongest lines from the optical and infrared line samples. The black crosses mark the equivalent widths as computed by direct 1D NLTE synthesis, while the orange line shows the result from interpolation between 1D NLTE syntheses at C abundances of 8.03, 8.43 and 8.83 using precomputed departure coefficients determined at the central value. The grey lines shows the resulting error in abundance space.}
    \label{fig:Interpolation}
\end{figure}

\begin{figure*}
    \centering
    \includegraphics[width=\linewidth]{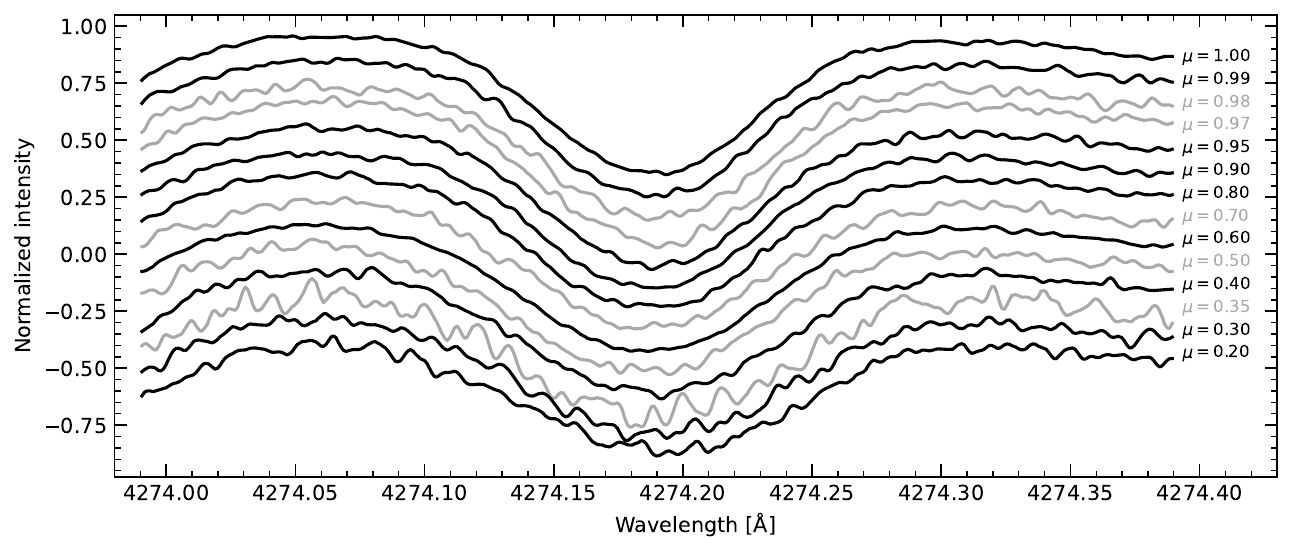}
    \caption{IAG observations of optical A-X transition at \SI{4274.19}{\angstrom} at various $\mu$-angles \protect\citep{EllwarthSchafer2023}. The angles starting from $\mu=0.99$ have been offset by 0.1 on the y-axis with respect to one another. The grey observations have been omitted in the final abundance analysis as these are of comparatively poor SNR.}
    \label{fig:IAG_CLV_4274}
\end{figure*}

\begin{figure*}
    \centering
    \includegraphics[width=\linewidth]{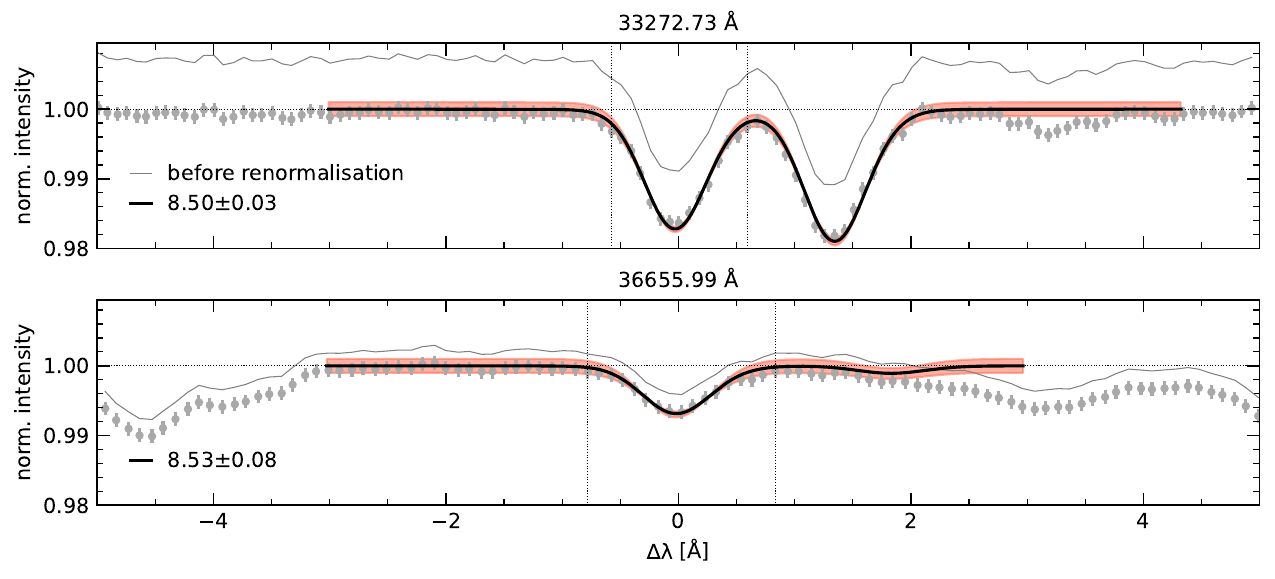}
    \caption{Impact of continuum normalisation on derived abundances for infra-red lines. The Spacelab-3 \texttt{ATMOS} observations after renormalisation by SUPPNet are shown as grey markers with errorbars. The thin grey line shows the original atlas. The black line shows the best-fit synthetic spectrum with the corresponding 3D non-LTE C abundance and associated uncertainty in the legend. The uncertainty in the abundance is determined by fitting the model to slightly offset copies of the observed spectrum. The resulting fits are framed by the red shaded area.}
    \label{fig:IR_fits3}
\end{figure*}

\begin{figure}
    \centering
    \includegraphics[width=\linewidth]{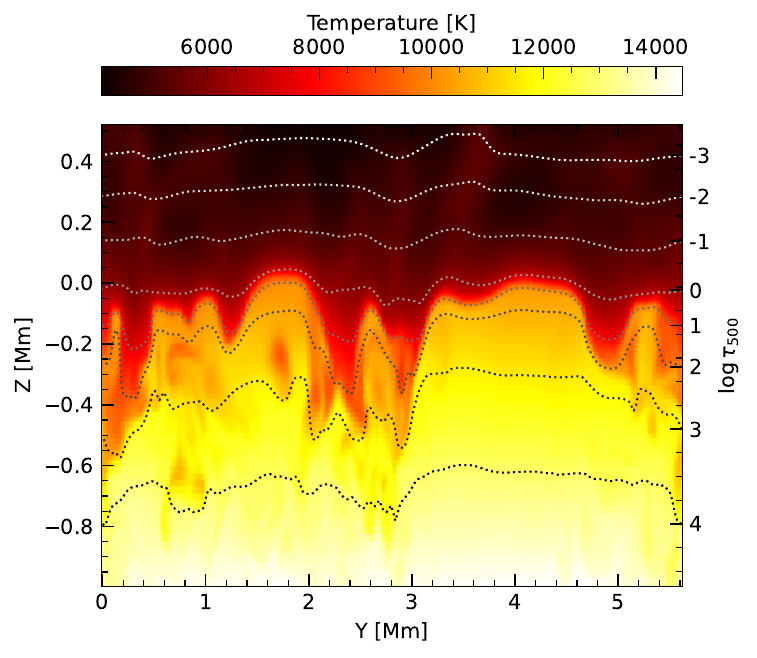}
    \caption{Temperature slice in YZ plane of one solar \Must snapshot. Here we have zoomed in around the optical surface and not shown the full extent of the snapshot. The full snapshot size is $6\times6\times2.75$ Mm.}
    \label{fig:temp2d}
\end{figure}

\begin{figure}
    \centering
    \includegraphics[width=\linewidth]{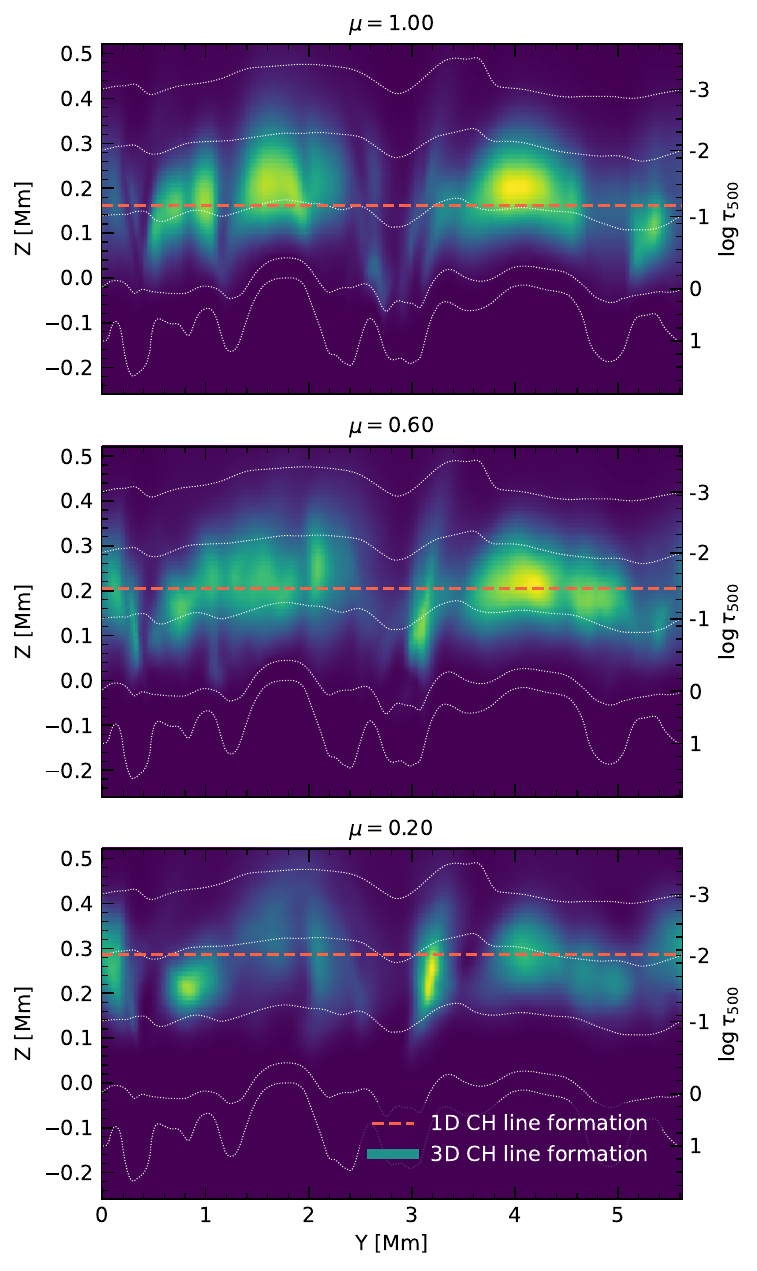}
    \caption{Intensity depression contribution of the optical \SI{4218.71}{\angstrom} line for different viewing angles. Shown is a YZ slice through one solar \Must snapshot. The red dashed line corresponds to the line formation height above the optical surface at $\log\tau_{500}=0$ with the maximal contribution in the 1D \Marcs model. The corresponding temperatures are shown in Fig. \ref{fig:temp2d}.}
    \label{fig:cntrbf2d}
\end{figure}

\begin{figure}
    \centering
    \includegraphics[width=\linewidth]{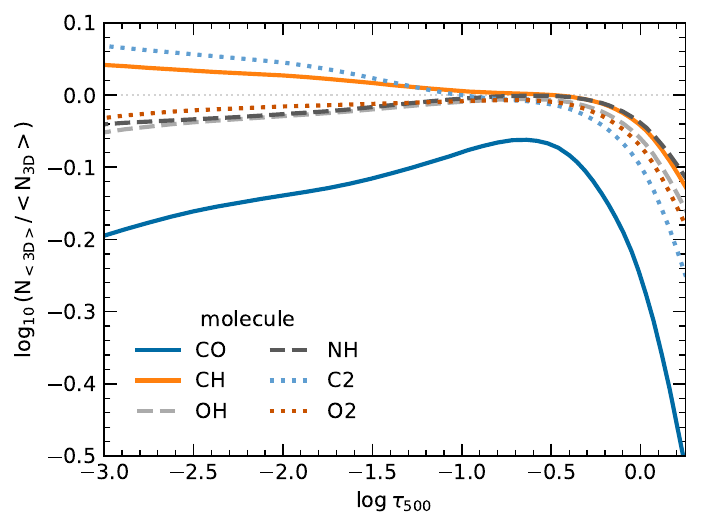}
    \caption{Ratio of molecular number densities for different molecules in an <3D> \Must snapshot over the average molecular number density of the full 3D snapshot as a function of optical depth. CH and C2 are the only molecules where using a <3D> atmosphere overestimates their average number density, thereby defying the prediction of \protect\cite{UitenbroekCriscuoli2011}. See section \ref{sec:molecules} for more details.}
    \label{fig:molecules}
\end{figure}

\subsection{Updates to the CH model molecule} \label{sec:updates}
All collision rates in PH23 were computed using an incorrect molecular weight for CH, leading to an overestimation of the collision rates. By adopting the NIST value of \SI{13.02}{u} for the molecular weight, we found that NLTE effects in the optical diagnostic lines increased by +0.02 dex. To improve precision, we now apply all collisional recipes directly at each atmospheric point, rather than linearly interpolating from a precomputed temperature grid, as was done in PH23. Since linear interpolation led to an overestimation of the rates, this adjustment resulted in an additional upward revision of the NLTE corrections by +0.02 dex.

However, the calculations in PH23 were performed using the \Multi code \citep{Carlsson1986}, which requires the dissociated "continuum" state to have the highest energy among all states in the model atom/molecule. Due to this limitation, the authors replaced the dissociation threshold of \SI{3.47}{\eV} with \SI{5.16}{\eV}. This incorrect dissociation energy only affected the calculation of collisional dissociation rates. The \cite{Seaton1962} and \cite{Drawin1969} rates include a factor of $\exp(-E_{ij}/(k_BT))$, where $E_{ij}$ is the energy difference between the lower and upper states. Consequently, the collisional dissociation rates were significantly underestimated, leading to an overall overestimation of the NLTE abundance corrections in PH23.

In this work, we have removed this limitation in our custom version of \MultiD, allowing unbound states to have energies between bound states of the same species. This enabled us to use the correct dissociation threshold of \SI{3.47}{\eV}. Despite the upward correction due to the revised molecular weight and the elimination of interpolation in the collision rates, this final adjustment had the most significant impact, ultimately reducing the derived abundance corrections to an absolute value of +0.01 dex on average. This underscores the strong influence that collisional transition rates have on departures from LTE.

\subsection{Comparison of the full vs. reduced model} \label{sec:reduced}
\begin{figure}
    \centering
    \includegraphics[width=\linewidth]{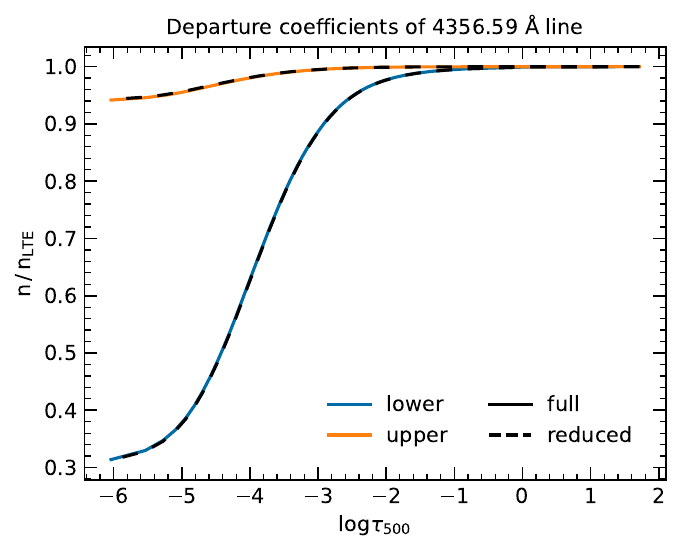}
    \includegraphics[width=\linewidth]{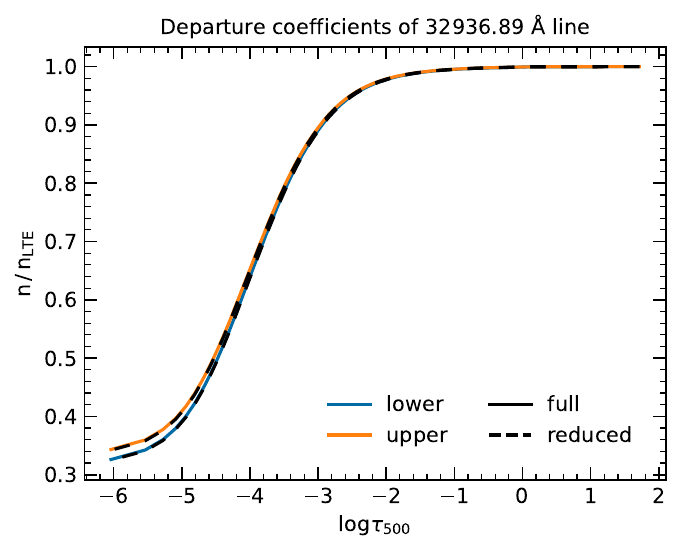}
    \caption{1D NLTE departure coefficients of the upper and lower energy states as a function of optical depth. Upper panel: A-X transition at \SI{4356.59}{\angstrom}.
    Lower panel: infrared rotation-vibration transition at \SI{32936.89}{\angstrom}. The coloured lines are the results based on the full model atom with 1981 levels. Overplotted as solid dashed lines are the results from the reduced model atom.}
    \label{fig:Departures1D}
\end{figure}

The resulting departure coefficients of the full and the reduced models are nearly indistinguishable. Fig. \ref{fig:Departures1D} shows 1D NLTE departure coefficients of the lower and upper energy states for two lines, namely one A-X transition and one infrared rotation-vibration transition. The lower state is underpopulated with respect to LTE populations in both cases. This is a result of efficient photo-ionisation in the upper parts of the atmosphere. The upper states of the A-X transitions (including the G-band) are generally underpopulated if the upper state is below the dissociation threshold or overpopulated if it lies above. In all cases the reduced model atom reproduces the behaviour from the full model atom. Both models converge below a 0.1\% change in the level populations within 4 iterations. Running the full model atom takes roughly 12 times as long.

\subsection{Optical lines: Equivalent width error analysis} \label{sec:EWerror}
We generally employ \texttt{Scipy}'s \texttt{curve\_fit} method for fitting models to observations. \texttt{curve\_fit} uses a non-linear least-squares fit and provides estimates of the statistical uncertainties in the fitted parameters. We determine equivalent widths (EW) by fitting Gaussian profiles to the observations after renormalising the spectral region of $\pm1.5$\AA{} around each diagnostic line. Due to the low SNR of 10-50 of the observations and surrounding blending lines, there is a significant uncertainty associated with the continuum placement. We quantify the systematic uncertainty in the continuum fit $\sigma^{\rm obs}_{\rm sys}$ by computing the standard deviation of the residuals between the observations and the 3D NLTE model where the normalised intensity is above $0.8$. In order to estimate the statistical uncertainty $\sigma^{\rm obs}_{\rm stat}$ we apply a lowpass filter with a cutoff frequency of \SI{15}{\per\angstrom} to the observations, removing high frequency noise and essentially smoothing the data. Then we take $\sigma^{\rm obs}_{\rm stat}$ to be the standard deviation of the residuals between the smoothed and unsmoothed observations. This leaves us with a normalised wavelength region and its statistical and systematic uncertainty, which is assumed to be the same for all frequency points.
We parametrise the Gaussian profile as
\begin{equation}
    I(\lambda) = 1 - \mathcal{A} \exp \left(- \frac{(\lambda - \lambda_0)^2}{2 \Delta^2} \right),
\end{equation}
where $\mathcal{A}$ is the amplitude and $\Delta$ is the width of the Gaussian. The EW in the spectrum follows from
\begin{equation}
    W = \sqrt{2\pi} \mathcal{A} \Delta.
\end{equation}
\texttt{curve\_fit} yields the statistical uncertainties on the amplitude $\sigma^{A}_{\rm stat}$ and standard deviation $\sigma^{\Delta}_{\rm stat}$ of the profile, which we convert into a statistical uncertainty on the EW
\begin{equation}
    \sigma^{\rm W}_{\rm stat} = \sqrt{2\pi} \sqrt{(\mathcal{A} \sigma^{\Delta}_{\rm stat})^2 + (\Delta \sigma^{\mathcal{A}}_{\rm stat})^2}
\end{equation}
The statistical uncertainty $\sigma^{\rm obs}_{\rm stat}$ is taken into account by \texttt{curve\_fit}, but we need to compute the systematic uncertainty on EW $\sigma^W_{\rm sys}$ manually. We do so by fitting another Gaussian to the same data, but minimally offsetting the continuum. This allows us to calculate the numerical derivative of the EW as a function of the continuum position  $\partial W / \partial C$. We can estimate $\sigma^W_{\rm sys}$ to be
\begin{equation}
    \sigma^W_{\rm sys} = \sigma^{\rm obs}_{\rm sys} \frac{\partial W}{\partial C}.
\end{equation}
and finally we combine the statistical and systematic uncertainties via

\begin{equation}
    \sigma^W_{\rm tot} = \sqrt{(\sigma^W_{\rm stat})^2 + (\sigma^W_{\rm sys})^2}.
\end{equation}

\subsection{Optical lines: Abundance error analysis} \label{sec:ABerror}
We determine best fit abundances by fitting the model EWs to the measured EWs at all angles simultaneously. Each EW measurement has an associated  uncertainty $\sigma^W_{\rm tot}$ which is mostly due to the uncertain continuum placement. The continuum placement is affected by the low SNR and surrounding blending lines. Due to the presence of the same line blends we consider the uncertainty in the continuum placement highly correlated between the different viewing angles. In fact, we assume the shared systematic uncertainty to be equal to the statistical uncertainty, hence: $\sigma^W_{\rm sys}=\sigma^W_{\rm stat}=\sigma^W_{\rm tot}/\sqrt{2}$. Here, we define $\sigma^W_{\rm sys}$ as the systematic uncertainty between individual EW measurements, in contrast to the last section, where it represented the uncertainty on the EW due to systematic uncertainties in the intensities at different frequencies. Once again \texttt{curve\_fit} provides us with a statistical uncertainty on the fitted abundance $\sigma^A_{\rm stat}$, based on the statistical uncertainty of the EW measurements. The dominant systematic uncertainty on the abundance fit $\sigma^A_{\rm sys}$, however, we estimate manually by computing the numerical derivative of the fitted abundance with respect to a small systematic EW offset. 

\subsection{Infra-red lines: Fitting procedure and error analysis}
\label{sec:IRfitting}
The infrared vibration-rotation lines shown in Figures~\ref{fig:IR_fits1} and~\ref{fig:IR_fits2} were fitted using the following procedure:

First, we selected a region around each line center to serve as the fitting window. To suppress high-frequency noise and enable numerical differentiation, a lowpass filter with a cutoff frequency of \SI{1.5}{\per\angstrom} was applied to the observed spectra. Next, we trimmed each region to the nearest surrounding maxima. Once this was done, we identified the nearest inflection points on either side of each line center. This allowed us to divide each line wing into two subregions:
\begin{itemize}
    \item Region 1: From the line center to the first inflection point.
    \item Region 2: From the inflection point to the nearest maximum.
\end{itemize}

Region 1 was always included in the fitting. Region 2 was included only if two criteria were met: (1) no additional inflection points were present in Region 2, and (2) the intensity difference between the line core and the nearest maximum covered at least $7/8$ of the line’s total amplitude. These conditions were designed to exclude any nearby blended lines from the fitting window.

Synthetic spectra were then convolved to match the resolving power of the observations using a Gaussian kernel, with Doppler broadening defined by
$
\Delta v = c / R \approx c \, \Delta\lambda / \lambda_0.
$
Finally, a $\chi^2$-fit was performed within the defined fitting windows. The free parameters in the fit were the C abundance and a small wavelength shift. For the 1D model fits, we also included an additional Gaussian broadening term.

The uncertainty in the carbon abundance was determined as follows:

The SUPPNet normalization provides wavelength-dependent errors on the continuum. We estimated the continuum uncertainty as the mean of these values within the fitting window, but enforced a minimum uncertainty of 0.1\%. Then we repeated the $\chi^2$-fit twice on the observed spectrum raised and lowered by the given uncertainty. From the three fits we construct a parabolic dependence of the abundance on the continuum level, which is finally differentiated to obtain an error on the best-fit abundance.

\begin{figure*}
    \centering
    \includegraphics[width=\linewidth]{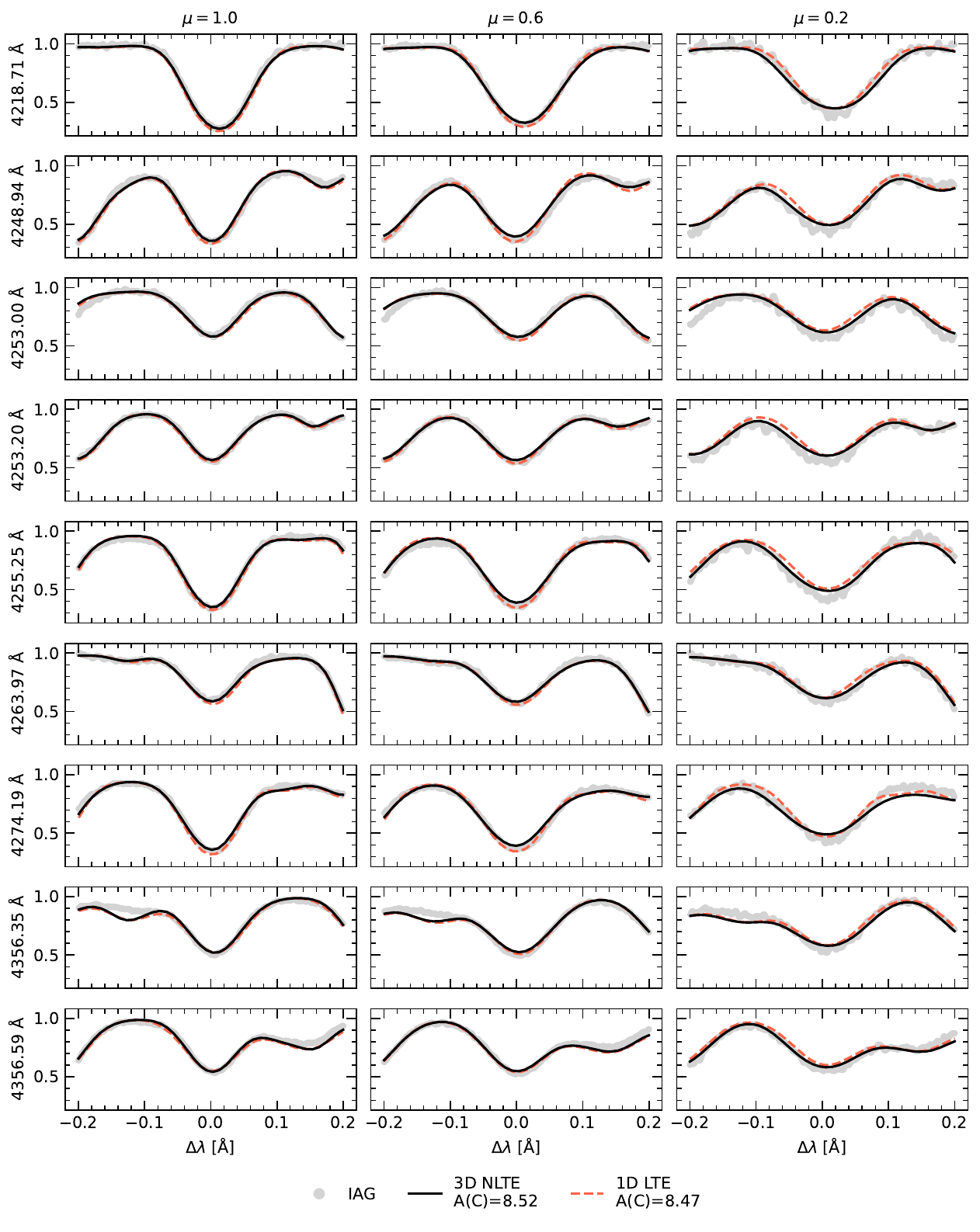}
    \caption{IAG spectra for $\mu=1.0,0.6,0.4$ of diagnostic A-X transitions. Overplotted are the 3D NLTE (\Must) and 1D LTE (\Marcs) line profiles of the best-fit abundances determined by matching all angles and A-X lines simultaneously. The 1D LTE profiles tend to be too strong at $\mu=1.0$ (disk-center) and too weak at $\mu=0.2$ (limb).}
    \label{fig:CH_AX_fit}
\end{figure*}

\begin{figure*}
    \centering
    \includegraphics[width=\linewidth]{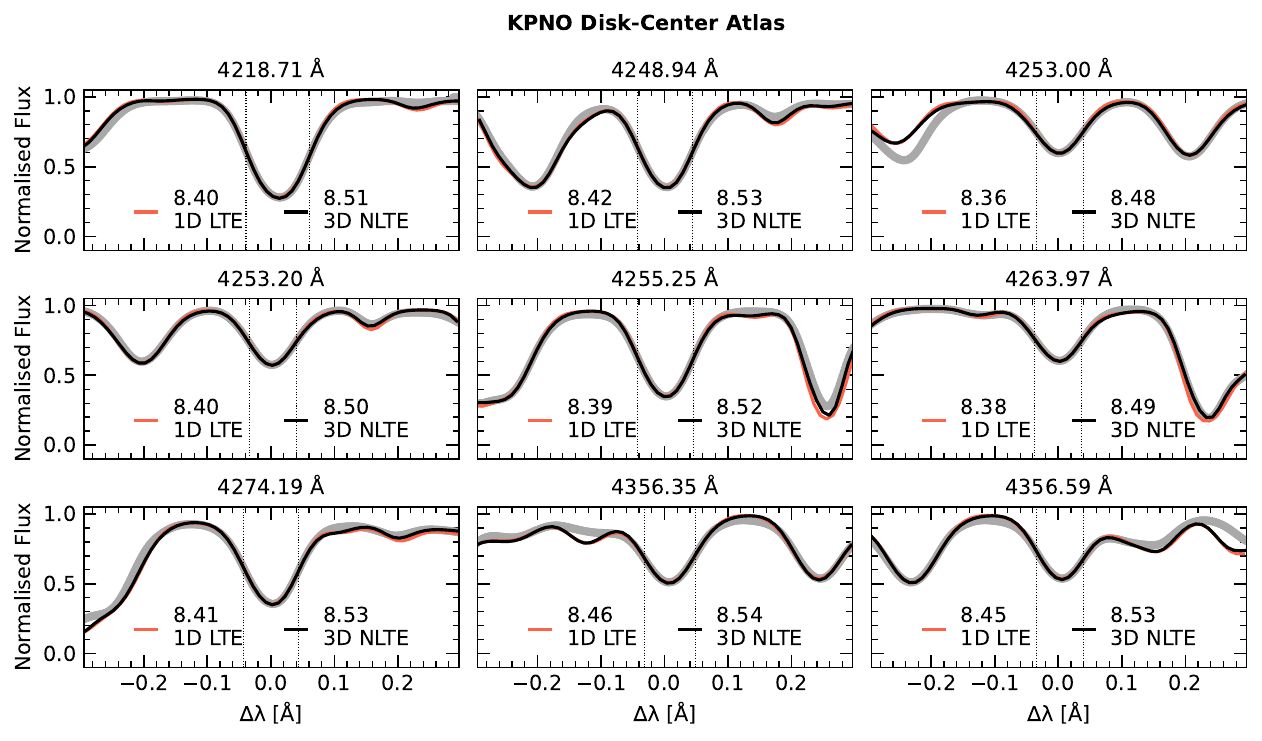}
    \includegraphics[width=\linewidth]{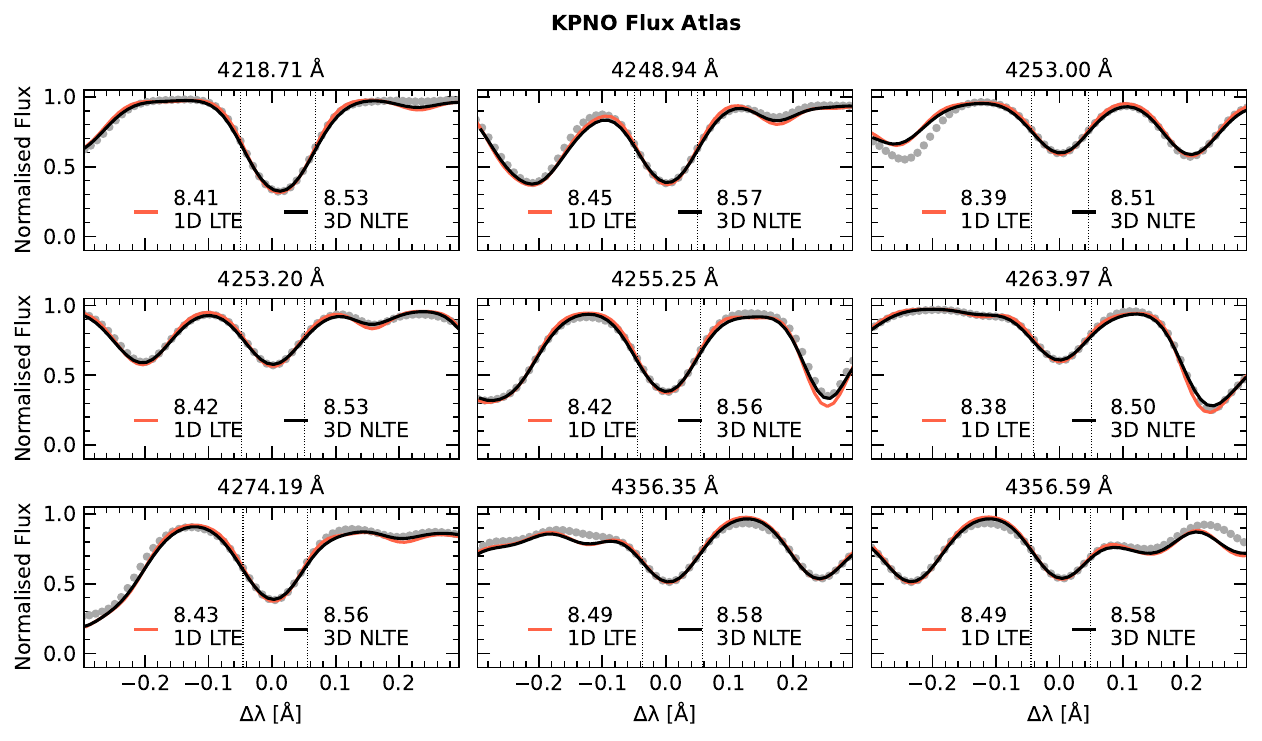}
    \caption{1D LTE (\Marcs) and 3D NLTE (\Must) best-fits of optical diagnostic A-X transitions to the KPNO disk-center \protect\citep{brault1987spectral} and flux \protect\citep{KuruczFurenlid1984} atlases. The legend labels show the corresponding C abundance. The vertical lines frame the windows used in the fitting-procedure.}
    \label{fig:CH_AX_KPNO}
\end{figure*}

\begin{figure*}
    \centering
    \includegraphics[width=\linewidth]{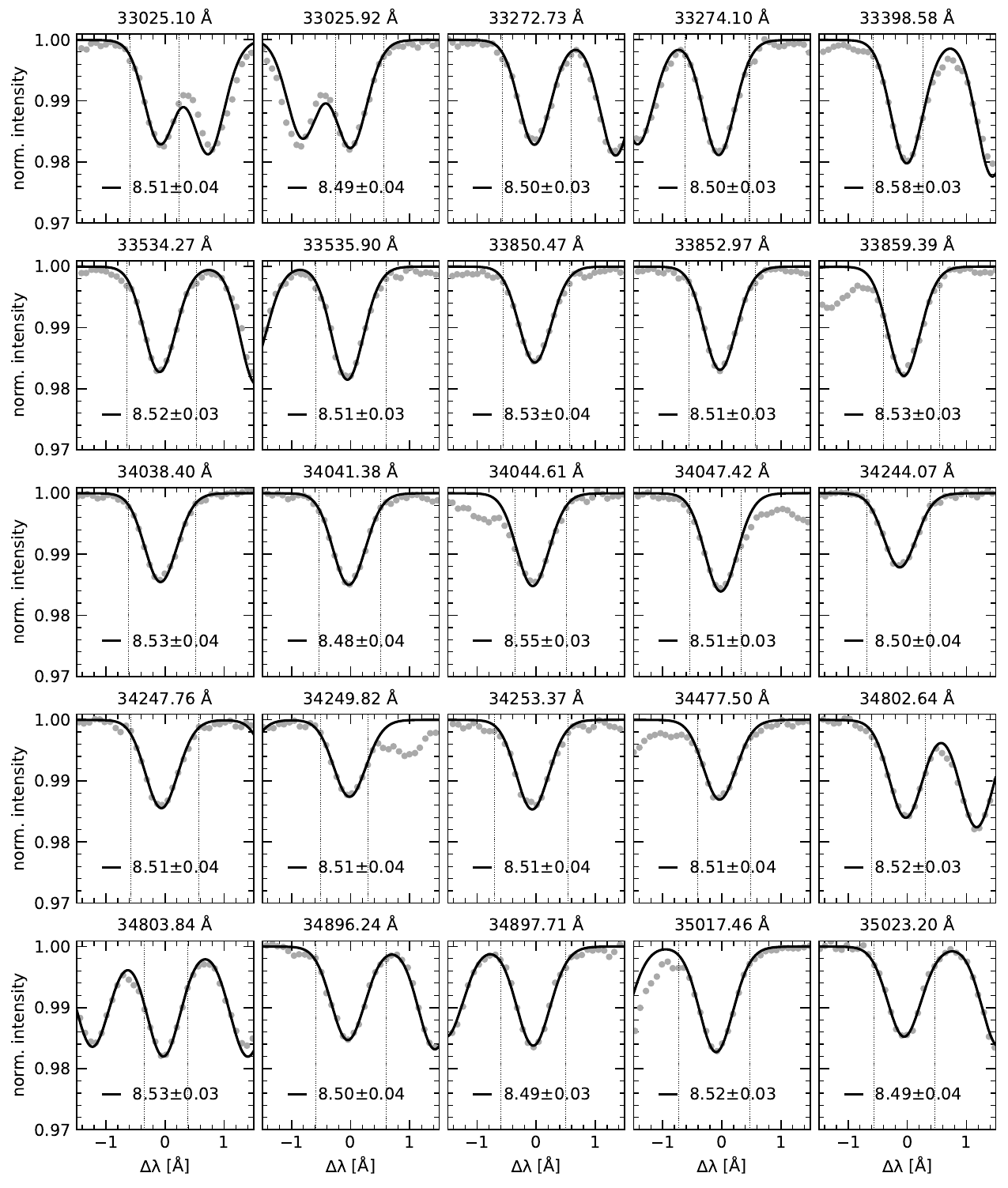}
    \caption{Spacelab-3 \texttt{ATMOS} atlas observations \protect\citep{FarmerNorton1989} and 3D NLTE (\Must) line profile fits of infra-red vibration-rotation lines. The legend labels show the best-fit C abundance. The associated uncertainty is dominated by the uncertainty in the continuum placement. The vertical lines frame the windows used in the fitting-procedure.}
    \label{fig:IR_fits1}
\end{figure*}

\begin{figure*}
    \centering
    \includegraphics[width=\linewidth]{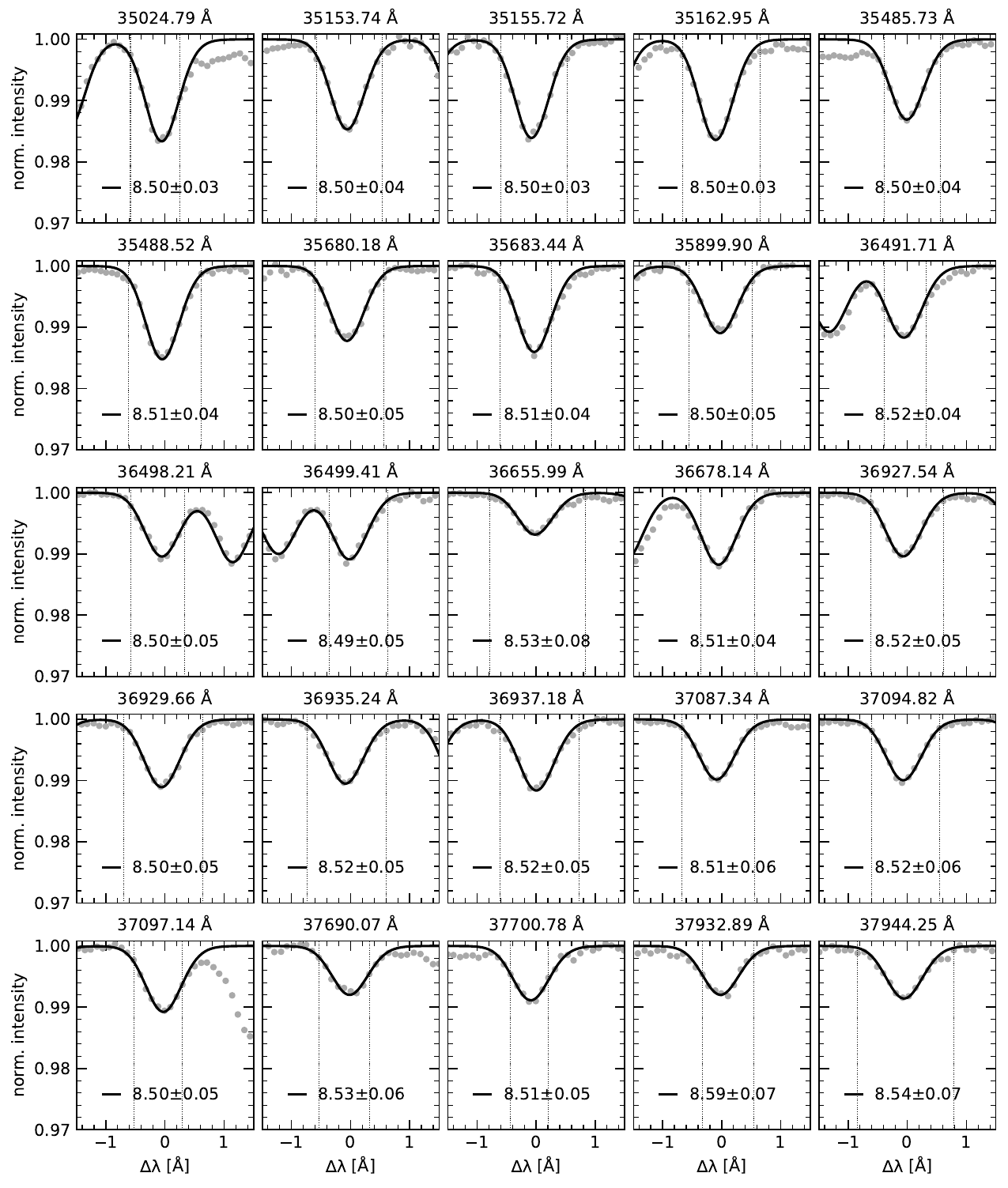}
    \caption{Continuation of Fig. \ref{fig:IR_fits1}.}
    \label{fig:IR_fits2}
\end{figure*}


\bsp	
\label{lastpage}
\end{document}